\documentclass[fleqn,usenatbib]{mnras}

\usepackage{newtxtext,newtxmath}

\usepackage[T1]{fontenc}
\usepackage{ae,aecompl}

\usepackage{graphicx}	
\usepackage{amsmath}	
\usepackage{amssymb}	
\usepackage[referable]{threeparttablex}
\usepackage{float}
\usepackage[T1]{fontenc}

\title[RAGN quenching neighbouring galaxies]{POSSIBLE EVIDENCE OF THE RADIO AGN QUENCHING OF NEIGHBOURING GALAXIES AT $z\sim1$}

\author[L Shen]{Lu Shen$^{1},$\thanks{Contact e-mail: \href{lushen@ucdavis.edu}{lushen@ucdavis.edu}}
Adam R. Tomczak$^{1}$,
Brian C. Lemaux$^{1}$,
Debora Pelliccia$^{1}$,
\newauthor{Lori M. Lubin$^{1}$,
Neal A. Miller$^{2}$,
Serena Perrotta$^{3}$,
Christopher D. Fassnacht$^{1}$,}
\newauthor{Robert H. Becker$^{1}$,
Roy R. Gal$^{4}$, 
Po-Feng. Wu$^{5}$, 
Gordon Squires$^{6}$,}
\\
$^{1}$Physics Department, University of California, Davis, One Shields Avenue, Davis, CA 95616, USA \\
$^{2}$Stevenson University, Department of Mathematics and Physics, 1525 Greenspring Valley Road, Stevenson, MD, 21153, USA\\
$^{3}$Department of Physics and Astronomy, University of California, Riverside, 900 University Ave, Riverside, CA 92521\\
$^{4}$University of Hawai'i, Institute for Astronomy, 2680 Woodlawn Drive, Honolulu, HI 96822, USA\\
$^{5}$Max-Planck Institut f\"{u}r Astronomie, K\"{o}nigstuhl 17, D-69117, Heidelberg, Germany\\
$^{6}$Spitzer Science centre, California Institute of Technology, M/S 220-6, 1200 E. California Blvd., Pasadena, CA, 91125, USA
}

\date{Accepted XXX. Received YYY; in original form ZZZ}

\pubyear{2018}

\begin{document}
\label{firstpage}
\pagerange{\pageref{firstpage}--\pageref{lastpage}}
\maketitle

\begin{abstract}
Using 57 Radio Active Galactic nuclei (RAGN) at 0.55 $\leq$ z $\leq$ 1.3 drawn from five fields of the Observations of Redshift Evolution in Large Scale Environments (ORELSE) survey, we study the effect of injection of energy from outbursts of RAGN on their spectroscopically-confirmed neighbouring galaxies (SNGs). 
We observe an elevated fraction of quenched neighbours ($\mathrm{f_q}$) within 500 kpc projected radius of RAGN in the most dense local environments compared to those of non-RAGN control samples matched to the RAGN population in colour, stellar mass, and local environment at 2$\sigma$ significance. 
Further analyses show that there are offsets at similar significance between $\mathrm{f_q}$s of RAGN-SNGs and the appropriate control samples for galaxies specifically in cluster environments and those hosted by most massive cluster galaxies, which tentatively suggests that some negative feedback from the RAGN is occurring in these dense environments. In addition, we find that the median radio power of RAGN increases with increasing local overdensity, an effect which may lend itself to the quenching of neighbouring galaxies. Furthermore, we find that, in the highest local overdensities, the $\mathrm{f_q}$ of the sub-sample of lower stellar mass RAGN-SNGs is larger than that of the higher stellar mass RAGN-SNGs sub-sample, which indicates a more pronounced effect from RAGN on lower stellar mass galaxies.
We propose a scenario in which RAGN residing within clusters might heat the intracluster medium (ICM) affecting both in situ star formation and any inflowing gas that remains in their neighbouring galaxies. 
\end{abstract}



\section{Introduction} 
On galaxy scales, radio active galactic nuclei (RAGN) deposit most of their energy and momentum into the interstellar, circumgalactic, or intergalactic medium via high-velocity jets, which may be responsible for the quenching of the star formation in the host galaxy or helping to maintain its quiescence (see \citealp{Fabian2012} for a review). However, it is not yet clear whether RAGN can have a significant effect on neighbouring galaxies. 

In principle, there exist many reasons to suggest that RAGN might affect other galaxies in their vicinity. 
Radio observations have revealed that powerful radio jets, originating from the center of the RAGN, extend for kiloparsecs or even megaparsecs beyond the host galaxy (see review paper by \citealp{McNamara2012}). Many X-ray observations have revealed cavities, bubbles and shocks in the hot intracluster medium (ICM) of some clusters, coincident with the lobes of the radio sources (e.g. \citealp{Boehringer1993, McNamara2000, Tremblay2010}). 
Moreover, RAGN hosted by the brightest cluster galaxies (BCGs) in the local Universe have been shown to release enough mechanical energy to heat the surrounding ICM and prevent rapid cooling in the centers of clusters (e.g., \citealp{Birzan2004, Best2007}). 

There are several channels that enable RAGN to efficiently interact with the ICM, such as displacing gas, driving shocks, or transporting low entropy gas and heavy elements outward from cluster cores. 
In the inner region ($\sim$30 kpc) of a cluster, RAGN heating via energetic outbursts released from RAGN drive shocks, which boost the entropy level and lift the temperature along the direction of the outbursts. 
In the outer region ($\sim$300 kpc and larger), hot and overpressurized bubbles released by RAGN produce a weak shock that heat the surroundings, as observed by Deep Chandra images of Hydra A \citep{Nulsen2005} and MS0735+7421 \citep{McNamara2005}. 
Furthermore, studies of hydrodynamical simulations show that multiple cycles of RAGN activity in galaxies both central to and interspersed throughout the cluster act as heating agents of the ICM (e.g. \citealp{DallaVecchia2004, Bruggen2005, Nusser2006}). 
\citet{Voit2005b} simulated the RAGN heating based on observed core gas entropy profiles and suggested that multiple cycles of RAGN outbursts incrementally heat and increase entropy by several keV cm$^2$ at $\sim$1 Mpc. 
These observations and simulations indicate RAGN are important components of the heating in the ICM, leaving open the possibility that such heating might be involved in the prevention of new episodes of star formation in both the host galaxies and its neighbours. 

However, observational evidence of such feedback is scarce, as the few studies that have observationally investigated such feedback have found minimal or no affect by RAGN on their surrounding galaxies. 
\citet{Shabala2011} studied photometrically selected neighbouring galaxies in two subsamples of RAGN in local groups and clusters, classified by radio morphology into $\sim$70 Fanaroff-Riley class II (FR-II) sources (edge-dominated radio emission) and 21 FR-I sources (core-dominated radio emission). They found that neighbouring galaxies in the projected jet paths of FR-II RAGN are redder than neighbouring galaxies outside the jet path, suggesting the radio jets might be capable of quenching neighbouring galaxies that reside in the jet path, but no such trend was found for the satellites of FR-I RAGN. 
Later, \citet{Pace2014} studied neighbouring galaxies within 100 kpc of a larger ($\sim$7000) sample of FR-I and FR-II RAGN selected from the SDSS and NVSS+FIRST surveys at z $\le$ 0.3. This population was compared to neighbouring galaxies of a control sample of non-radio emitting galaxies that were matched in redshift, colour, luminosity, axis ratio, and environment. They found that RAGN have more red and quenched neighbouring galaxies compared to that of the control sample in all types of  environments to which they were able to compare (i.e., field, cluster galaxy and brightest cluster center). However, they stated that this is largely due to more neighbouring galaxies around RAGN. 

While few studies have attempted to investigate such RAGN feedback, none of them have done so at intermediate redshifts (z$\sim$1) and across a wide range of environments. 
Therefore, we will approach this question by studying a population of RAGN neighbouring galaxies and those galaxies in proximity to a matched control sample of non-RAGN galaxies selected in various environments at intermediate redshifts (0.55 $\leq$ z $\leq$ 1.3) across five fields in the Observation of Redshift Evolution in Large Scale Environments (ORELSE; ~\citealp{Lubin2009}) survey. 
The ORELSE survey is a systematic search for large-scale structures (LSSs) in $\sim$0.25-0.5 deg$^2 $ around an original sample of 18 galaxy clusters in a redshift range of 0.55 $\leq$ z $\leq$ 1.3. 
This survey targets galaxy populations over a wide range of local and global environments. 
We briefly introduce the observational data and reduction in section~\ref{sec:obs}. In section~\ref{sec:samples}, we describe our method for selecting the RAGN and control sample, as well as their neighbouring galaxies. In Section~\ref{sec:result}, we show our results on the quiescent fractions of neighbouring galaxies of RAGN and analyses on its potential cause. In Section~\ref{sec:discussion}, we then discuss the results and propose a scenario to explain them.
Throughout this paper all magnitudes, including those in the infrared (IR), are presented in the AB system \citep{Oke1983, Fukugita1996}. All distances are quoted in proper units. We adopt a concordance $\Lambda$CDM cosmology with $\mathrm{H_0 = 70~km~s^{-1} Mpc^{-1}}$, $\Omega_{\Lambda} = 0.73$, and $\mathrm{\Omega_{M} = 0.27}$, and a Chabrier stellar initial mass function (IMF;~\citealp{Chabrier2003}).
\vspace{-5mm}

\section{DATA and METHODS}
\label{sec:obs}
In this section, we introduce the observational data and describe the reduction of those data, the method adopted to estimate local and global environment measurements, and the calculation of the quiescent fraction ($\mathrm{f_q}$).  
In this paper, we use five fields (SC1604, SG0023, SC1324, RXJ1757, RXJ1821) from the ORELSE survey which have fully reduced radio catalogues and accompanying photometric and spectroscopic catalogues. These observations span 7$\sim$15 Mpc in the plane of the sky and encompass 11 clusters and 17 groups, spanning a total (dynamical) mass range of $10^{12.8} \mathrm{M}_\odot$ to $10^{15.1} \mathrm{M}_\odot$. See \citet{Shen2017} and \citet{Rumbaugh2012} for more details on these five fields. 

\begin{table*}
	\caption{Number of Spectroscopic objects, radio galaxies, RAGN and SNGs in each field }
	\label{tab:fields}
	\begin{threeparttable}
	    \begin{tabular}{lccccccccc}
		    \hline
		    \hline
	    	Field & R.A.\tnote{1} & Decl.\tnote{1} & Num. of &  Spec\tnote{2}  & Radio-spec\tnote{3} & Num. of & Num. of & RAGN Size\tnote{6}\\
	    	& (J2000) & (J2000) & Clusters(Groups)\tnote{4}& Galaxies & galaxies & RAGN & RAGN-SNG\tnote{5} & (arcsec)\\
		    \hline
		    SC1604 & 16:04:15 & +43:21:37 & 6(4)& 960(583) & 108 & 27 & 72 & $4.8^{+1.4}_{-0.3}$\\
         	    SG0023 & 00:23:52 & +04:22:51 & 0(6) & 655(420) & 46 & 9 & 12 &  $5.5^{+2.0}_{-0.2}$\\
		    SC1324 & 13:24:35 & +30:18:57 & 3(5) & 673(530) & 38 & 8 & 20 & $6.8^{+1.6}_{-1.8}$\\
		    RXJ1757 & 17:57:19.4 & +66:31:29 &1(1) & 257(186) & 26 & 4 & 5 & $5.2^{+0.2}_{-0.1}$\\
		    RXJ1821 & 18:21:32.4 & +68:27:56 &2(0)& 231(200) & 32 & 9 & 39 & $5.2^{+0.6}_{-0.3}$\\ 
		    \hline
	    \end{tabular}
	    \begin{tablenotes}
		\item[1] Coordinates for SC1604, SG0023, SC1324 are the median of central positions of clusters/groups, while RXJ1757 and RXJ1821 are given as the centroid of the peak of diffuse X-ray emission associated with the respective cluster.  
        \item[2] Number of secure spectroscopically confirmed galaxies in the redshift range 0.55 $\leq$ z $\leq$ 1.3, with $18.5 \le i'/z \le 24.5$ (for more details see Section~\ref{sec:specobs}). The number in parentheses denotes the subset of these galaxies with $\mathrm{M_* \geq 10^{10}M_\odot}$ and $\mathrm{M_{NUV}-M{r} \geq 2}$ limit. (for more details see Section~\ref{sec:Ngals}).
		\item[3] Number of radio sources matched to all spectroscopic confirmed galaxies.
		\item[4] Number of clusters (groups) in each field. 
		\item[5] Number of spectroscopically-confirmed neighbouring galaxies (SNGs) around RAGN sample in each field. See Section \ref{sec:Ngals} for details on SNG selection criteria. 
		\item[6] The median value of RAGN size and the 1$\sigma$ scatter (i.e. 16\% and 84\% values) in each field. 
	    \end{tablenotes}
   \end{threeparttable}
\end{table*}

\subsection{Imaging and Photometry}
\label{sec:photobs}
Comprehensive photometric catalogues are constructed for all five fields. We summarize the available optical and near-infrared (NIR) observations and the reduction process here. See \citet{Tomczak2017} and Tomczak et al. \textit{(submitted)} for specific details and additional information regarding the photometry used in this study. 

Optical imaging was taken with the Large Format Camera (LFC;~\citealp{Simcoe2000}) on the Palomar 5-m telescope, using Sloan Digital Sky Survey (SDSS, \citealp{Doi2010})-like r$^{\prime}$, i$^{\prime}$ and z$^{\prime}$ filters, reduced in the Image Reduction and Analysis Facility (IRAF,~\citealp{Tody1993}), following the method in~\citet{Gal2008}. We also use R${^+}$, R$_{c}$, I${^+}$, I$_{c}$, Z${^+}$ and Y band optical imaging from Suprime-Cam~\citep{Miyazaki2002} on the Subaru 8-m telescope, reduced with the \texttt{SDFRED2} pipeline~\citep{Ouchi2004} supplemented by several routines from Traitement \'El\'ementaire R\'eduction et Analyse des PIXels (\texttt{TERAPIX})\footnote{http://terapix.iap.fr}. 
Some J and K band data were taken with the United Kingdom Infra-Red Telescope Wide-Field Camera (WFCAM;~\citealp{Hewetti2006}) mounted on the United Kingdom Infrared Telescope (UKIRT) and was reduced using the standard UKIRT processing pipeline courtesy of the Cambridge Astronomy Survey Unit\footnote{http://casu.ast.cam.ac.uk/surveys-projects/wfcam/technical}. 
Additionally, J and Ks band imaging was taken using the Canada-France-Hawaii Telescope Wide-field InfraRed Camera (WIRCam;~\citealp{Puget2004}) mounted on the Canada-France-Hawai'i Telescope (CFHT) and was reduced through the I'iwi preprocessing routines and \texttt{TERAPIX}. Infrared imaging at 3.6, 4.5, 5.8, and 8.0 $\mu$m (5.8 and 8.0 $\mu$m only available for SC1604) was taken using the Spitzer telescope Infrared Array Camera (IRAC;~\citealp{Fazio2004}). The basic calibrated data (cBCD) images provided by the Spitzer Heritage Archive were reduced using the MOsaicker and Point source EXtractor (MOPEX; \citealp{Makovoz2005}) package augmented by several custom Interactive Data Language (IDL) scripts written by J. Surace. 

Photometry was obtained by running Source Extractor (SExtractor;~\citealp{Bertin1996}) on point spread function (PSF)-matched images convolved to the image with the worst seeing. Magnitudes were extracted in fixed circular apertures to ensure that the measured colours of galaxies are unbiased by different image quality from image to image. Also, the package T-PHOT~\citep{Merlin2015} was used for Spitzer/IRAC images, due to the large point spread function of these data that can blend profiles of nearby sources together and contaminate simple aperture flux measurements. 

Spectral Energy Distribution (SED) fitting is performed in a two-stage process. First, we used the Easy and Accurate Redshifts from Yale (\textsc{EAZY};~\citealp{Brammer2008}) code to estimate photometric redshifts ($z_{phot}$) for galaxies that lacked spectroscopic redshifts. Rest-frame colours are also derived in this step using the best-fit $z_{phot}$ ($z_{spec}$ when available) which are used to classify galaxies as star-forming or quiescent (see Section \ref{sec:qf}). 
In the second step, we used the Fitting and Assessment of Synthetic Templates (\textsc{FAST}; ~\citealp{Kriek2009}) code to estimate stellar masses as well as other properties of the stellar populations of galaxies. In brief, \textsc{FAST} creates a multi-dimensional cube of model fluxes from a provided stellar population synthesis (SPS) library. Each object in the photometric catalogue is fit by every model in this cube by minimizing $\chi^2$ for each model and adopted the model of the lowest minimum $\chi^2$ as the best-fit. For this we make used of the SPS library presented by \citet{Bruzual2003}, assuming a Chabrier (2003) stellar initial mass function, allowing for dust attenuation following the \citet{Calzetti2000} extinction law. See Section 2.3 of \citet{Tomczak2017} for a more thorough description of these procedures and assumptions. It is important to note that we have tested our analysis using stellar masses derived from a smaller stellar population synthesis (SPS) modeling parameter space and find that our results of RAGN classification and stellar mass distributions are unaffected (see \citealt{Shen2017} for a description of this parameter space). Based on a visual inspection of the best-fit SEDs of all RAGN, we find a good agreement between the models and the observed photomerty, which gives us confidence that the reported stellar mass and dust extinctions are representative of host galaxies despite the presence of non-stellar emission. To further underscore this point, the median $\chi^2$ of the best-fit SEDs of all RAGN is 1.19 compared to 0.97 for the median $\chi^2$ value of the full photometric catalogue (with good use flag).

The precision of the photometric redshifts were estimated from fitting a Gaussian to the distribution of $\mathrm{(z_{spec} - z_{phot})/(1 + z_{spec})}$ measurements in the range 0.5 $\le$ z $\le$ 1.2. We find values of $\sigma_{\Delta z/(1+z)}$ ranging from 2.9 - 3.2\% for all five fields to a limit of $i' \le 24.5$ and a catastrophic outlier rate ($\mathrm{\Delta z/(1 + z) > 0.15}$) of 4.8-9.5\%. 
Additionally, photometric sources are limited to 18.5 $\leq$ I $\leq$ 24.5. These are the same limits that we will apply to the spectroscopic sample (see Section~\ref{sec:specobs}) as a compromise between maximizing the sample size and the spectroscopic completeness. For structures at z$>$0.95 the Z band was used instead with the same limits. 

\subsection{Spectroscopy}
\label{sec:specobs}
Spectroscopic targets were selected based on the optical imaging in the $r'$, $i'$, and $z'$ from LFC imaging following the methods in~\citet{Lubin2009}.  In brief, the spectroscopic targeting scheme employed a series of colour and magnitude cuts that are applied to maximize the number of targets with a high likelihood of being on the cluster/group red sequence at the presumed redshift of the LSS in each field (i.e., priority 1 targets). However, the fraction of priority 1 targets which entered into our final sample ranged from 10\% to 45\% across all ORELSE fields, a fraction which tended to vary strongly with the density of spectroscopic sampling per field (see \citealp{Tomczak2017} for more discussion). 
In addition, for certain masks we prioritized X-ray and radio detected objects.  
The optical spectroscopy was primarily taken with the DEep Imaging and Multi-Object Spectrograph (DEIMOS;~\citealp{Faber2003}) on the Keck II 10m telescope and reduced using a modified version of the Deep Evolutionary Exploratory Probe 2 (DEEP2, \citealt{Davis2003,Newman2013}) pipeline. See Lemaux et al.\ (\textit{submitted}) for details on the modifications to the pipeline.
A few additional redshifts from the Low Resolution Imaging Spectrometer (LRIS; \citealp{Oke1995}) were added for SC1604, SG0023 and RXJ1821 (see~\citealp{Oke1998, Gal2004, Gioia2004}). For more details on the spectroscopic observations in these fields see \citet{Lemaux2012,Lemaux2017} and \citet{Rumbaugh2017}. 

Spectroscopic redshifts of these targets were extracted and assessed using the methods of~\citet{Newman2013}, while serendipious detections were added following the method described in~\citet{Lemaux2009}. 
Only galaxies with high quality redshifts (Q=3,4; see \citealt{Gal2008, Newman2013} for the meaning of these values) are used in this study. Spectroscopic samples are limited to the same $i'$/$z$ band limits ($18.5 \le i^\prime/z^\prime \le 24.5$) as applied to the photometric catalogues to keep these two catalogues consistent. 
We additionally apply $\mathrm{M_* \geq 10^{10} M_{\odot}}$ and $\mathrm{M_{NUV}-M_{r} \geq 2}$ cuts to the spectroscopically confirmed galaxy sample. These stellar mass and colour limits define where our spectroscopic sample is representative of the underlying photometric sample at 0.55 $\leq$ z $\leq$ 1.3 subject to the magnitude cuts above (see \citealt{Shen2017}, \citealt{Tomczak2017} and Lemaux et al. \textit{submitted.} for more details). This final spectroscopic sample consists of 1919 galaxies out of $\sim 2500$ galaxies with secure spectral redshift (i.e. Q=3,4), reliable photometry, and within the adopted rest-frame colour $\mathrm{M_{NUV}-M_{r}}$ and stellar mass range. The numbers in the samples for each field are listed in Table \ref{tab:fields}. In the reminder of this paper, we use this spectroscopic catalogue to construct the RAGN, control and their neighbouring galaxy samples.

\subsection{Radio Observations}
\label{sec:radioobs}
All five fields studied here were observed using the Very Large Array (VLA) 1.4GHz imaging in its B configuration, where the resulting FWHM resolution of the synthesized beam is about 5$^{\prime\prime}$ and has a circular field of view of $\sim31^\prime$ diameter centered on the optical images (i.e., the FWHM of the primary beam).
Net integration times were chosen to result in final $1\sigma$ sensitivities of about 10$\mu$Jy per beam for each field, a value which was approximately achieved for all five fields. 
The final images were then used to generate source catalogues. The NRAO's Astronomical Image Processing System (AIPS) task SAD created the initial catalogues by examining all possible sources having peak flux density greater than three times the local RMS noise. Peak flux density, integrated flux density, their associated  flux density errors ($\sigma$), and size are measured. We then instructed it to reject all structures for which the Gaussian fitted result has a peak below four times the local RMS noise. In the last step, we added those extended sources poorly fitted by Gaussians. 
We use the peak flux density unless the integrated flux is larger by more than $3\sigma$ compared to the peak flux for each individual source. 
The final radio catalogues contain sources above 4$\sigma$ and down to a flux density limit of about 30$\mu$Jy. For more details on the VLA reduction see \citet{Shen2017}.

\subsection{Environmental Measurements}
\label{sec:env}
There are two environment regimes: local environment which probes the current density field to which a galaxy is subject, and global environment which probes the time-averaged galaxy density to which a galaxy has been exposed.

\subsubsection{Local Environmental Density}
We adopt a local environment measurement using a Voronoi Monte-Carlo (VMC) algorithm which is described in full detail in \citet{Lemaux2017} and \citet{Tomczak2017}.  
In brief, 78 thin redshift slices ($\Delta\nu = \pm 1500~\mathrm{km/s}$) are constructed across the broad redshift range of 0.55 $\leq$ z $\leq$1.3, with adjacent slices overlapping by half depth of the slice. Spectroscopically-confirmed galaxies are then placed into these thin redshift slices. For each slice, photometric galaxies without a high quality $\mathrm{z_{spec}}$ (with a good use flag) have their original $\mathrm{z_{phot}}$ perturbed by an asymmetric Gaussian with a mean and dispersion set to the original $\mathrm{z_{phot}}$ and $\pm1\sigma$ uncertainty, respectively, and the objects whose new $\mathrm{z_{phot}}$ fall in the redshift slice are determined. 
Then a Voronoi tessellation is performed on the slice and is sampled by a two dimensional grid of 75 $\times$ 75 proper kpc pixels. Local density is defined as the inverse of the cell area multiplied by the square of the angular diameter distance. 
The final density map of the slice is computed by median combining the density values at for each pixel from 100 VMC realizations. 
The local overdensity value for each pixel point (i, j) is then computed as $\mathrm{log(1 + \delta_{gal}) = log(1 + (\Sigma_{VMC} - \tilde{\Sigma}_{VMC})/\tilde{\Sigma}_{VMC})}$, where $\mathrm{\tilde{\Sigma}_{VMC}}$ is the median $\mathrm{\Sigma_{VMC}}$ for all grid points over which the map is defined. Local overdensity rather than local density are adopted for mitigating issues of sample selection and differential bias on redshift. 

\subsubsection{Global Environmental Density}
To quantify the global environment, we adopt $\mathrm{R_{proj}/R_{200}}$ versus $\mathrm{|\Delta\nu|/\sigma_\nu}$~\citep{Carlberg1997, Balogh1999, Biviano2002, Haines2012, Noble2013, Noble2016}, defined by $\eta$ from \citet{Shen2017} as:  
\begin{equation}  
\label{eq:eta}
\mathrm{\eta =(R_{proj}/R_{200})\times(|\Delta\nu|/\sigma_\nu)}
\end{equation} 
where $\mathrm{R_{proj}}$ is the distance of each galaxy to each group/cluster center, $\mathrm{R_{200}}$ is the radius at which the cluster density is 200 times the critical density, $\Delta\nu$ is the difference between each galaxy velocity and the systemic velocity of the cluster, and $\sigma_\nu$ is the line-of-sight (LOS) velocity dispersion of the cluster member galaxies (see \citealt{Lemaux2012} for the computation of $\Delta\nu$ and $\sigma_\nu$). The cluster centers are obtained from the i$^\prime$-luminosity-weighted center of the members galaxies as described in \citet{Ascaso2014}. 
The value of $\eta$ for each galaxy is measured with respect to the closest cluster/group. To determine it, we first find all the clusters/groups that are within $\pm$6000 km s$^{-1}$ in velocity space of a given galaxy, then we compute $\mathrm{R_{proj}/R_{200}}$ from the galaxy to all the identified clusters and groups, and we select the one for which $\mathrm{R_{proj}/R_{200}}$ is the smallest as the parent cluster/group. If for a given galaxy no clusters/groups within $\pm$6000 km s$^{-1}$  are found, $\eta$ is computed with respect to all of those clusters/groups in the field and the one with the smallest value is associated with that galaxy. See \citet{Pelliccia2018} for more detail on this calculation.   
In this paper, when necessary we use $\eta \leq 2$ as the restriction for galaxies in cluster/group environments (see Section \ref{sec:test_cluster}). 
Note that while most massive clusters/groups are detected in our fields, many lower mass systems are missed (Hung et al. \textit{in prep.}). For this reason, the $\eta$ values calculated here are necessarily upper limits. 

\subsection{Classification of Star-forming and Quiescent Galaxies}
\label{sec:qf}
We adopt the rest-frame $\mathrm{M_{NUV} - M_{r}}$ versus $\mathrm{M_{r} - M_{J}}$ colour-colour diagram separations from \citet{Lemaux2014a}, based on the two-colour selection technique proposed by \citet{Ilbert2010}, to divide galaxies into two categories: quiescent and star-forming (SF). Specifically, galaxies at 0.55 $\leq$ z $\leq$ 1.0 with $\mathrm{M_{NUV} - M_r > 2.8(M_r - M_J) + 1.51}$ and $\mathrm{M_{NUV} - M_r > 3.75}$ and galaxies at 1.0 $<$ z $\leq$ 1.3 with $\mathrm{M_{NUV} - M_r > 2.8(M_r - M_J) + 1.36}$ and $\mathrm{M_{NUV} - M_r > 3.6}$ are considered quiescent. The fraction of quiescent galaxies ($\mathrm{f_q}$) for a galaxy population is calculated as the number of quiescent galaxies over the total number in full sample. Uncertainties in $f_q$ are derived from Poissonian statistics. 
As a reminder, we additionally applied stellar mass $\mathrm{M_* \geq 10^{10} M_{\odot}}$ and colour $\mathrm{M_{NUV}-M_{r} \geq 2}$ cuts to our entire secure spectroscopically confirmed galaxy sample (see Section \ref{sec:specobs}). 

\section{CONSTRUCTION of SAMPLES}
\label{sec:samples}
To search for possible evidence of RAGN influencing the star-forming properties of neighbouring galaxies, we first identify RAGN in our survey and construct a comparison sample of galaxies that are matched in nearly every relevant property (stellar mass, local environment, colour) as control samples. For the remainder of the paper, we refer to the RAGN and controls as ``centre''\footnote{Note that ``central'' has been used extensively to refer the most massive galaxies in the parent halo. We name the RAGN and control samples as ``centers'' to avoid confusion to readers.} galaxies. 
Then, we select the Spectroscopically-confirmed Neighbouring Galaxy (SNG) sample around the RAGN (RAGN-SNG) and control samples (control-SNG). We describe these selections below. 

\subsection{Radio AGN Sample}
\label{sec:agn_sample}

We perform a maximum likelihood ratio (LR) technique, following the precedures in Section 3.4 in \citet{Rumbaugh2012}, to search for optical counterparts to radio sources. 
In brief, a likelihood ratio is a quantity that estimates the excess likelihood that a given optical source is the genuine match to a given radio source relative to its arrangement arising by chance. The LR is given by the equation 
\begin{equation}
    LR_{i,j} = \frac{w_i exp(-r^2_{i,j}/2\sigma^2_j)}{\sigma^2_j}
	\label{eq:LR}
\end{equation}
Here, $r_{i,j}$ is the separation between objects i and j, $\sigma_j$ is the positional error of object j, where we use $1\arcsec$ as the positional error of all radio sources~\citep{Condon1997}, and $w_i = n(< m_i)^{-1/2}$ is the square root of the inverse of the number density of optical sources with magnitudes fainter than the observed $i'$ band magnitude. 
We then carried out a Monte Carlo (MC) simulation to estimate the probability that each optical counterpart is the true match using the LRs. The threshold for matching to a single or double objects is the same as that used in~\citet{Rumbaugh2012}, though in practice in this paper, only the highest probability matched optical counterpart was considered for each radio source. 
The optical matching is done to the overall photometric catalogues. We use a search radius of $1\arcsec$, aimed at being inclusive, i.e., not to miss any genuine matches due to instrumental/astrometric/astrophysical effects. 
We then search for extended morphology radio sources and radio doubles using a larger search radius following the same method used in Section 3.4 in \citet{Shen2017}. We add in one radio double in SC1324 in this step. 
In this paper, we focus on radio objects which have photometric counterparts with secure spectroscopically-confirmed redshifts and within the redshift range $0.55\leq z \leq1.3$. We refer to these galaxies as radio galaxies. The number of radio galaxies is listed in Table~\ref{tab:fields}. 

RAGN were then selected from this pool of radio galaxies following the two-stage radio classification presented in \citet{Shen2017}. First, all radio galaxies with log($\mathrm{L_{1.4GHz}) \geq 23.8}$ are immediately classifies as RAGN. Some radio galaxies below this luminosity threshold are also classified as RAGN if they are in the AGN region in the rest-frame $\mathrm{M_U - M_B}$ versus stellar mass normalized radio luminosity $\mathrm{L_{1.4GHz}/M_*(M_\odot)}$ (colour-SRL) diagram, as defined below: 
\begin{equation}\label{eq:classification}
  \begin{cases} 
     \mathrm{M_U - M_B} > 1.24  & \quad \text{if } \text{SRL} < 12.3 \\
     \mathrm{M_U - M_B} > -0.7\text{SRL} + 9.85  & \quad \text{if } 12.3 \leq \text{SRL} < 13.5 \\
     \mathrm{M_U - M_B} > 0.4  & \quad \text{if } \text{SRL} \leq 13.5
   \end{cases}
\end{equation}
where $M_U - M_B$ is the dust corrected colour. In order to correct for dust we use the value of the colour excess, E$_s$(B-V), as estimated for each galaxy from our SED fitting and apply a correction to each rest-frame absolute magnitude following the~\citet{Calzetti2000} reddening law. 
We note that the colour-SRL diagram is calibrated using radio galaxies at all redshifts, and no redshift dependence was found for the classification regions in this diagram. We obtain a total of 57 RAGN in the redshift range of $0.55 \leq z \leq 1.3$. 

\begin{figure}
    \includegraphics[width=\columnwidth]{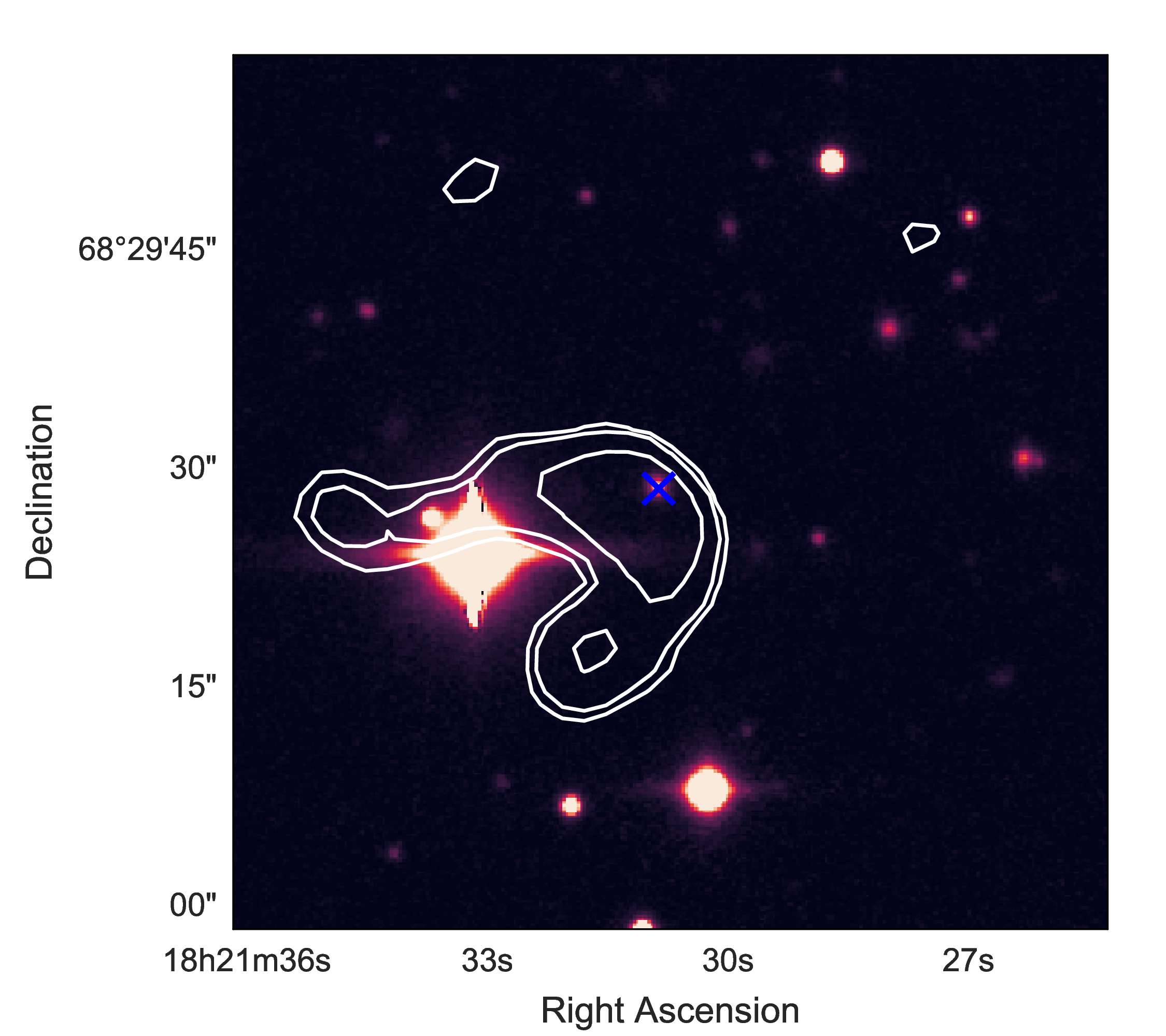}
    \vspace{-0.5cm}
    \caption{A cutout of the Y band image taken from the Suprime-Cam on the Subaru 8-m telescope centered on an extended radio active galactic nucleus (RAGN) in one of the ORELSE clusters in the RXJ1821 field. The image is displayed in color rather than grayscale purely for presentation purposes. Overlaid on the near infrared cutout are radio contours of three different levels: 3, 5, and 15 times the 9.33$\mu$Jy RMS of the radio image. The RAGN, which is also one of the MMCG is marked with a blue cross. The cutout is 1$\arcmin$ on each side, which translates to 474 kpc at $z=0.9$. This is only one source having extended radio profile and being hosted by an MMCG. }
    \label{fig:cutout}
\end{figure}

We manually examine the size of RAGN measured from the task SAD (see Section \ref{sec:radioobs}). We found that only two sources out of a total of 57 sources in the RAGN sample that were not well fit by this process. The rest of RAGN are well fit as point sources on the order of the beam size. For those two sources not well-fit by SAD, we estimate the circularized effective angular diameter of 15.2 and 13.1 arcsec, the lobe sizes from center to the local flux density peak are 19.5 and 18.0 arcsec, corresponding to 154 and 143 proper kpc at z=0.9, respectively. The median values and 1$\sigma$ scatters (i.e. 16\% and 84\% values) of RAGN sizes are listed in Table \ref{tab:fields}. One of the extended RAGN sources is also the Most Massive Cluster Galaxy (MMCG) in the most massive cluster in RXJ1821. The optical cutout centered on this extended RAGN with radio contours is shown in Figure \ref{fig:cutout}. See Section \ref{sec:test_cluster} for more discussion on this special case. 

\subsection{Control sample}
\label{sec:control}
To isolate the effect of RAGN on their neighbouring galaxies, we also select galaxies as a comparison sample from the spectroscopically confirmed galaxy samples which excludes RAGN and their neighbouring galaxies within 1 Mpc (see section \ref{sec:Ngals}, referred to as ``non-RAGN spectroscopic sample''). This comparison sample is identified using a 3D matching algorithm, following the \citet{Shen2017} method, which ensures that the distributions of M$_*$, rest-frame colour, and local environment of the control sample match closely to those of the RAGN sample. 
In brief, we split a 3D phase space ($\mathrm{M_{NUV} - M_{r}}$,  $\mathrm{log(M_*/M_\odot)}$, $\mathrm{log(1+\delta_{gal})}$) into $4 \times 4 \times 4$ boxes and calculate a 3D probability density map by taking a ratio of the number of galaxies from the RAGN sample in each box over the total number of RAGN. An individual "control" sample set is created by randomly sampling from the non-RAGN spectroscopic sample without replacement based on the 3D probability density map and having the same size of the RAGN sample. In order to explore the full breadth of possible outcomes for this comparison, we construct 100 such control samples. 

In \citet{Shen2017}, the matching was performed on two rest-frame colours ($\mathrm{M_{NUV} - M_{r}}$ and $\mathrm{M_{r} - M_{J}}$), but here we chose to match only on the former one because we have also added an environment parameter to the matching which decreases the number of RAGN and spectroscopic galaxies residing in each multi-dimensional box.  
We chose $\mathrm{M_{NUV} - M_{r}}$ as the rest-frame colour control parameter because it brackets the 4000\AA~break. In addition, when comparing to the other parameter ($\mathrm{M_{r} - M_{J}}$) in the colour-colour diagram, $\mathrm{M_{NUV} - M_{r}}$ distinguishes red and blue galaxies better than the latter and, thus, has more discriminatory power for measurements related to the age of the luminosity-weighted dominant stellar population of galaxies. These measurements are strongly linked to the classification of a galaxy as quiescent or star forming. 

\begin{figure}
    \includegraphics[width=\columnwidth]{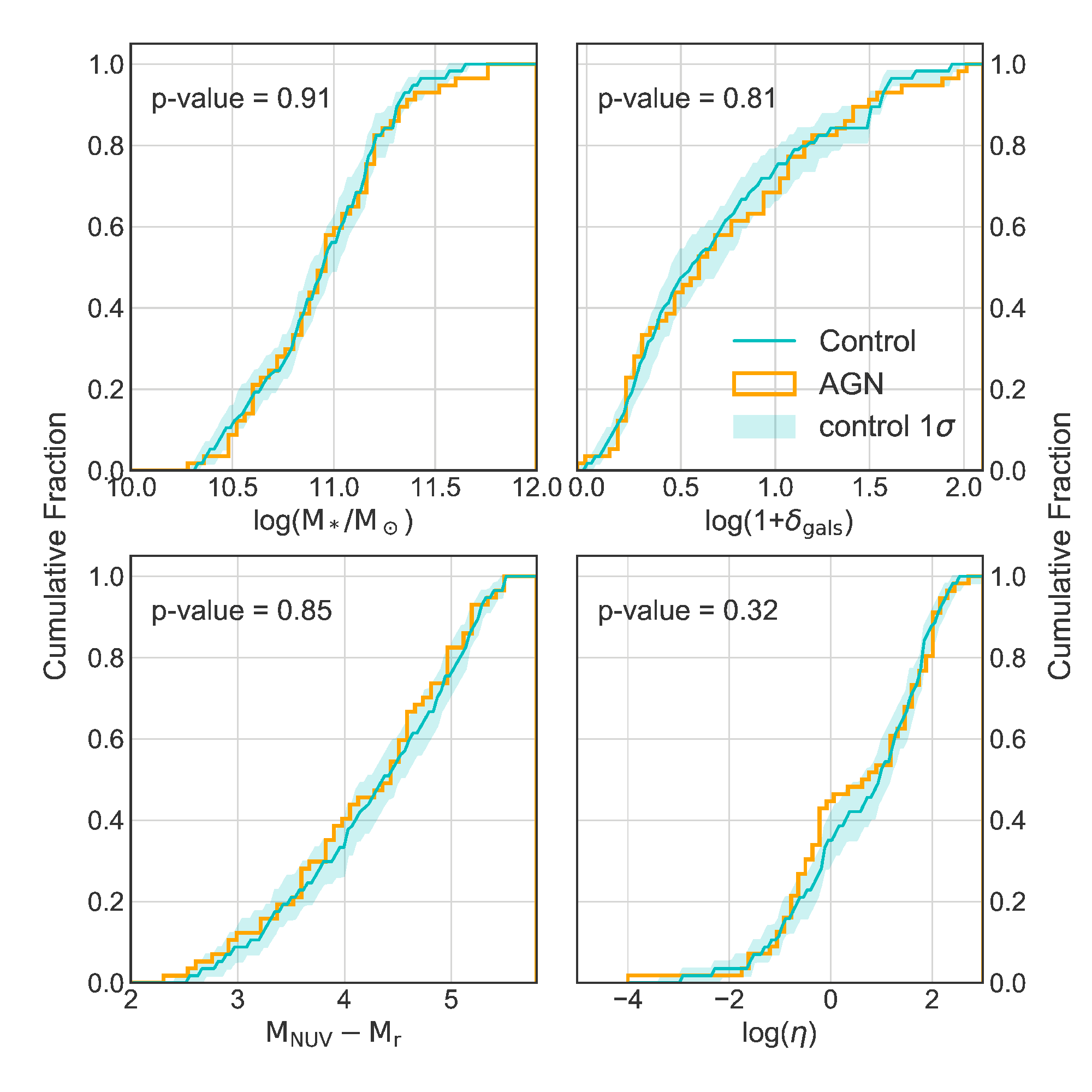}
    \caption{CDFs of RAGN (orange lines) and control samples (blue lines) on stellar mass, colour ($\mathrm{M_{NUV}-M_{r}}$), local overdensity ($\mathrm{log(1+\delta_{gal})}$), and the global environment (log($\eta$)). Note that we did not explicitly match the two samples in the last parameter. The blue lines are median values of the 100 control samples. The shaded regions are the 1$\sigma$ spread (i.e., the 16\% and 84\% values) of the 100 control samples. K-S tests run on these two distributions for each parameter yield no significant differences, with p-values listed in plots. }
    \label{fig:control}
\end{figure}

In figure~\ref{fig:control}, we show the comparison of RAGN and control samples in stellar mass, colour, local overdensity, and global environment. Note that we did not explicitly match on the last property. 
We used the median value of 100 cumulative distribution functions (CDFs) of control samples to compare against the RAGN sample.
The 1$\sigma$ lower and upper errors of each property of control samples were calculated as the 16\% and 84\% values, respectively, on the 100 samples. 
We use a Kolmogorov-Smirnov statistic  (K-S) test between the distributions of the RAGN and control samples to assess their similarity. The K-S test returns an effective probability (p-value) that the two observed distributions are drawn from the same parent distribution. We adopt p < 0.05 as a threshold to indicate that the RAGN and control samples are sufficiently different to each other. 
The p-values are 0.91, 0.85, 0.81 and 0.32 for the $\mathrm{log(M_*/M_\odot)}$, $\mathrm{M_{NUV}-M_{r}}$, $\mathrm{log(1+\delta_{gal})}$ and log($\eta$) comparisons, respectively. In all cases, the K-S p-value was not sufficiently small to confirm that the control samples are drawn from an underlying population that was significantly different in its colour, stellar mass, and environmental properties from that of the control sample. 

\subsection{Neighbouring galaxies}
\label{sec:Ngals}
\begin{figure}
    \includegraphics[width=\columnwidth]{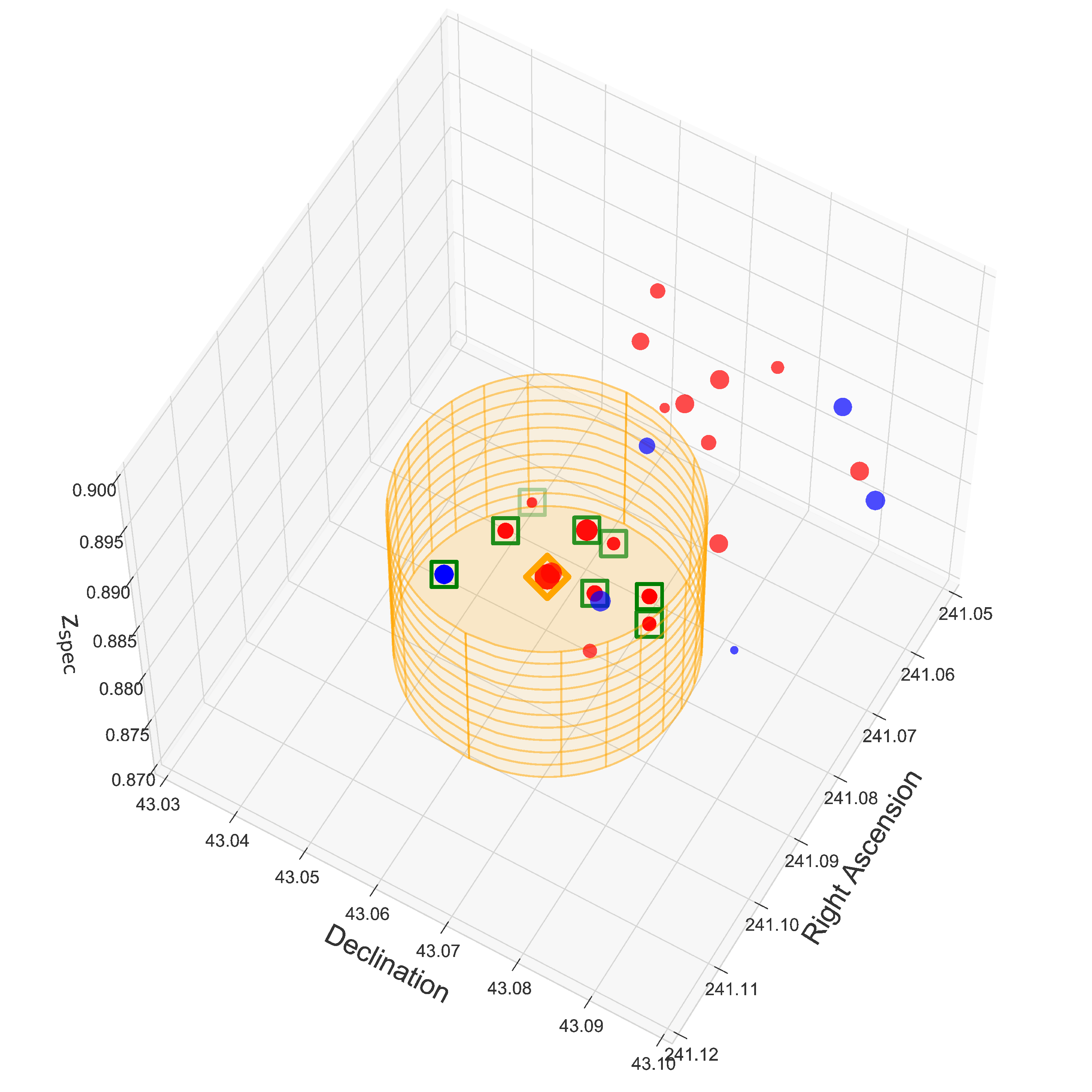}
    \caption{\textit{Top:} Demonstration of the selection of neighbouring galaxies with a RAGN as a centre. The orange cylinder represent the selection volumes centered on each centre having radii of $\mathrm{R_{proj}}$ = 500 kpc and depths of $\Delta$z = $\pm$1000 km/s. Dots are spectroscopically-confirmed galaxies in the 3-dimensional space, colour-coded in red and blue for quiescent and star forming galaxies, respectively. The size of dots are scaled by their stellar mass. The selected neighbouring galaxies are boxed by open squares, and the centre RAGN is boxed by an open diamond.}
    \label{fig:neighbouring}
\end{figure}

In the final step, we selected the spectroscopically-confirmed neighbouring galaxies (SNGs) around the RAGN and control samples. 
SNGs are selected in cylindrical volumes centered on each centre having radii of $\mathrm{R_{proj}}$ = 500 kpc and depths of $\Delta$z = $\pm$1000 km/s, as shown in Figure~\ref{fig:neighbouring}.
The projected distance threshold was chosen to include all possible nearby galaxies that might reasonably be affected by proximity to the radio activity (e.g.,~\citealp{Tremblay2010}). 
The velocity requirement was chosen to maximize the purity and completeness of neighbouring galaxies, based on the typical galaxy velocity dispersion ($\sigma_v \sim 500-1000$ km s$^{-1}$) along the line of sight of clusters/groups in these five fields (see~\citealt{Lemaux2012} on how $\sigma_v$ is calculated). 
In section \ref{sec:test_radius}, we will further discuss how varying the radial cut affects the results. 
For cases where SNGs are selected around two or more RAGN/control galaxies, we assigned these SNGs to the nearest projected RAGN/control galaxy. 

\begin{table*}
	\caption{Number of centre and their SNGs in all tests}
	\label{tab:numbers}
	\begin{threeparttable}
	    \begin{tabular}{l | c | cccccc | c | c} 
		    \hline
		    \hline
	    	Samples & Total\tnote{1}  & $\mathrm{log(1+ \delta_{gal})}$ &  $0.4 \mathrm{log(1+ \delta_{gal})}$ & $0.8 \mathrm{log(1+ \delta_{gal})}$ & $\mathrm{log(1+ \delta_{gal})}$ & Cluster & MMCGs\tnote{3} \\
		 & &  < 0.4 & < 0.8 &  < 1.2 & $ \geq 1.2$ & ($\eta \leq 2$) &  \\
		    \hline
		    RAGN & 57 & 20 & 14 & 12 & 11 & 24 & 11 \\
         	    RAGN-SNGs & 148 & 5 & 33 & 39 & 71 & 106 & 67\\
		    control\tnote{2} & 57 & 22$\pm$2 & 14$\pm$3 & 10$\pm$2 & 11$\pm$3 & 24 & 11\\
		    control-SNGs\tnote{2} &  112$\pm$19 & 5$\pm$3 & 27$\pm$10 & 30$\pm$11 & 48$\pm$17 & 73$\pm$12 & 56 \\
		    \hline
	    \end{tabular}
	    \begin{tablenotes}
		\item[1] Total number of centres and their SNGs across all four overdensity bins. 
        		\item[2] Numbers of control-SNGs are the median values of the 100 re-samplings, while uncertainties are 84\% and 16\% values if they exist. 
		\item[3] Numbers of most massive cluster galaxies (MMCGs) in each sample. See Section~\ref{sec:test_cluster} for definition of MMCGs.
	    \end{tablenotes}
   \end{threeparttable}
\end{table*}

\section{Results}
\label{sec:result}

\begin{figure}
    \includegraphics[width=\columnwidth]{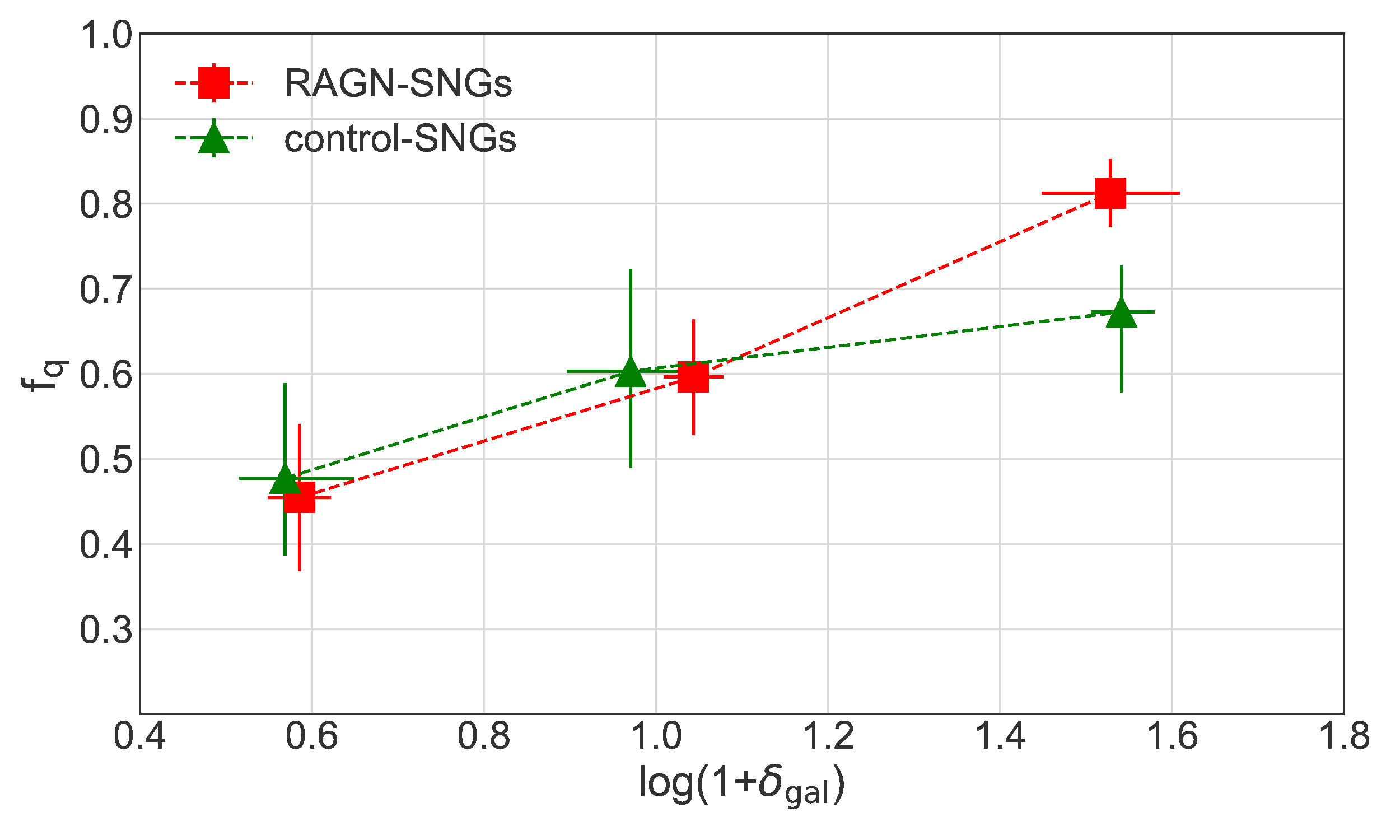}
    \caption{$\mathrm{f_q}$ for RAGN-/control-SNGs as a function of $\mathrm{log(1+\delta_{gal})}$ of the centre RAGN/control sample. Uncertainties on $\mathrm{f_q}$ are derived from Poissonian statistics. Uncertainties on $\mathrm{log(1+\delta_{gal})}$ of RAGN-SNGs are the $\mathrm{\sigma_{NMAD}/\sqrt{n-1}}$ (see Section \ref{sec:result}). The error bars on $\mathrm{f_q}$ and $\mathrm{log(1+\delta_{gal})}$ of control-SNGs are the asymmetric errors (i.e., 16\% and 84\% values of the median values of 100 re-samplings). Note that the lowest overdensity bin ($\mathrm{log(1+\delta_{gal})} < 0.4$) is removed because only 2 RAGN-SNGs are found in this overdensity bin; therefore, no comparison can be made in this overdensity bin.}
    \label{fig:result}
\end{figure}

We calculate the fractions of quiescent galaxies ($\mathrm{f_q}$s) for RAGN-SNGs and control-SNGs by binning the two samples into four local overdensity bins, adopting the $\mathrm{log(1+\delta_{gal})}$ values of the RAGN hosts and control galaxies in all cases. 
For control samples, we obtain the $\mathrm{f_q}$s of SNGs binned by $\mathrm{log(1+\delta_{gal})}$ of their centers in each control sample and use the median value to represent the true value of $\mathrm{f_q}$ for the overall control samples. The asymmetric errors of control sample are calculated as the 16\% and 84\% values on the 100 samples, which represent the variation of the control population. Uncertainties on the median $\mathrm{log(1+\delta_{gal})}$ of RAGN-SNGs are given by $\mathrm{\sigma_{NMAD}/\sqrt{n-1}}$ where $\mathrm{\sigma_{NMAD}}$ is the normalized median of the absolute deviations \citep{Hoaglin1983} and n is the number of the sample (see \citealp{Muller2000} for a discussion on adopting this type of estimate on the uncertainty on the median). Throughout the paper we conservatively adopt the $\mathrm{\sigma_{NMAD}/\sqrt{n-1}}$ as the formal uncertainty on $\mathrm{log(M_*/M_\odot)}$, $\mathrm{log(1+\delta_{gal})}$, log($\eta$) and z of RAGN/RAGN-SNGs, and 16\% and 84\% values of median values of the 100 re-samplings as the asymmetric error on those of control/control-SNGs. 

As shown in Figure~\ref{fig:result}, we find that the $\mathrm{f_q}$ of RAGN-SNGs (0.81$\pm0.04$) is marginally higher than that of control-SNGs (0.67$^{+0.06}_{-0.09}$)only in the highest overdensity bin ($\mathrm{1.2 \le log(1+\delta_{gal}) \leq 2.1}$). The difference between the two $\mathrm{f_q}$s is 0.14, a discrepancy significant at a $2\sigma$ level. This significance persists even if we match the median of $\mathrm{log(1+\delta_{gal})}$ of control-SNGs samples to that of the RAGN-SGNs sample, by removing the lowest local overdensity control-SNG in the highest overdensity bin. Additionally, this difference persists at the same level if we change the framework of our analysis to include different SED fitting parameters, binning, or colour/stellar mass cuts. 
To study the possible cause of this potential difference, we perform several diagnostic tests described below.  

\subsection{Assessing the Representativeness of the Control Sample}
\label{sec:test_comparison}

\begin{figure}
    \includegraphics[width=\columnwidth]{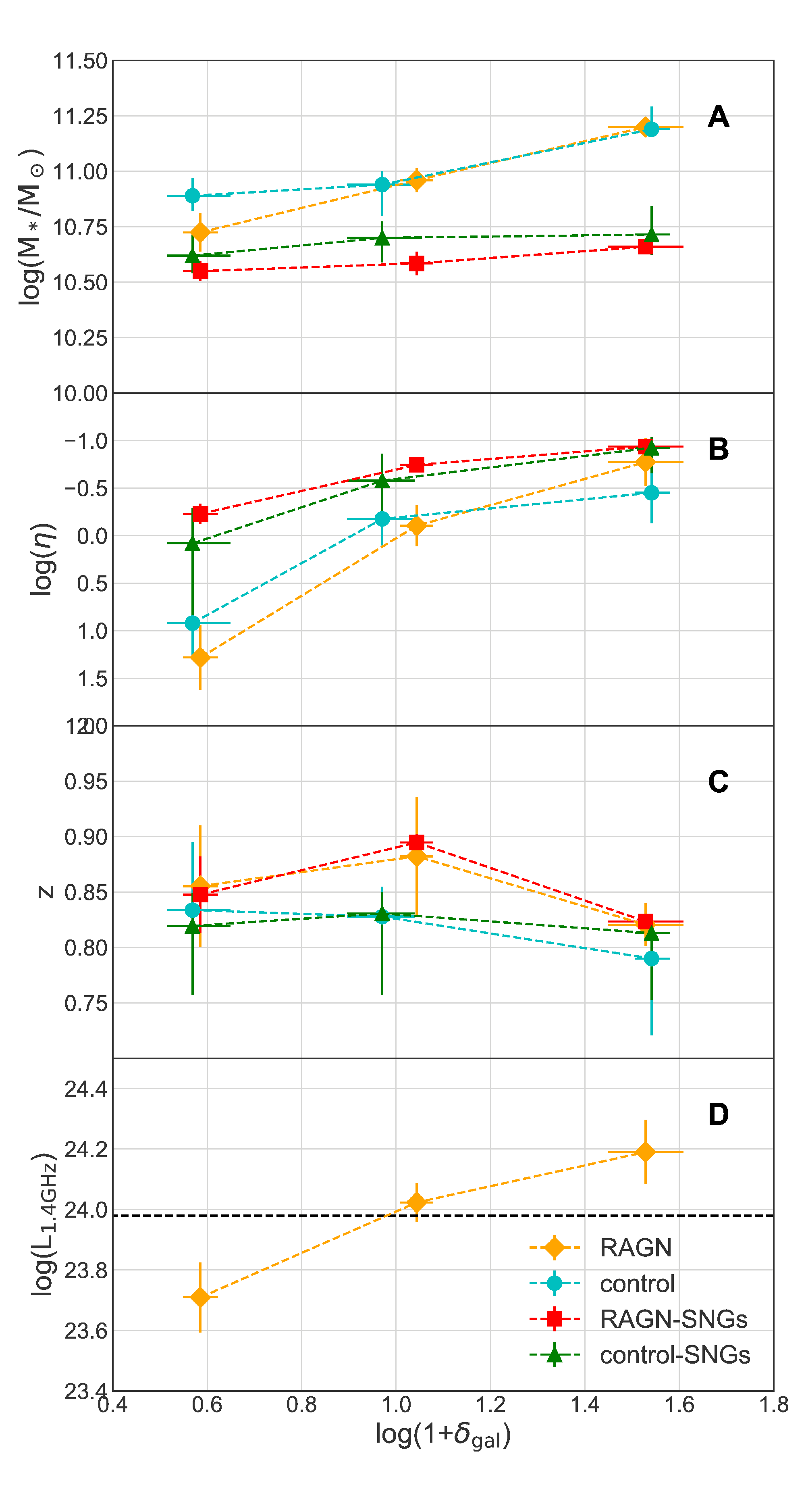}
    \vspace{-1cm}
    \caption{\textit{Panel A}: Stellar mass of RAGN (orange), control (cyan), RAGN-SNGs (red) and control-SNGs (green) as a function of the median $\mathrm{log(1+\delta_{gal})}$ of the centre RAGN/control sample. \textit{Panel B}: log($\eta$) of RAGN, control, RAGN-SNGs and control-SNGs as a function of the median $\mathrm{log(1+\delta_{gal})}$ of the centre RAGN/control sample. \textit{Panel C}: redshift of RAGN, control, RAGN-SNGs and control-SNGs as a function of the median $\mathrm{log(1+\delta_{gal})}$ of the centre RAGN/control sample. \textit{Panel D}: Radio luminosity at 1.4GHz of RAGN, binning by $\mathrm{log(1+\delta_{gal})}$. The median value ($\mathrm{log(L_{1.4GHz})} = 23.98 \pm 0.11$) of the full RAGN sample is shown by the black dashed line. Uncertainties on $\mathrm{log(1+\delta_{gal})}$, $\mathrm{log(M_*/M_\odot)}$, log($\eta$) and z of RAGN/RAGN-SNGs are the $\mathrm{\sigma_{NMAD}/\sqrt{n-1}}$ (see Section \ref{sec:test_comparison}). The error bars on $\mathrm{log(1+\delta_{gal})}$, $\mathrm{log(M_*/M_\odot)}$, log($\eta$) and z of control/control-SNGs are the asymmetric errors (i.e., 16\% and 84\% values of the median values of 100 re-samplings). }
    \label{fig:validation}
    \vspace{-0.4cm}
\end{figure}

First, we want to investigate whether this is a real physical effect caused by the RAGN or just systematic differences between the RAGN-SNGs and control-SNGs. Studies have shown that the quiescent fraction depends on galaxy stellar mass, redshift (e.g.~\citealt{Muzzin2013, Muzzin2014, Tomczak2014} and Lemaux et al. \textit{submitted}), and its presence in a cluster/group or field environment (e.g.~\citealt{Hansen2009, Cooke2016}). 
To exclude these factors, we run comparisons between these parameters of the RAGN-SNGs and control-SNGs.

As shown in panel A of Figure~\ref{fig:validation}, the binned median $\mathrm{log(M_*/M_\odot)}$ values of RAGN-SNGs versus control-SNGs overlap across the three overdensity bins, indicating no stellar mass dependence in the neighbouring galaxies driving the $\mathrm{f_q}$. 
We also perform a K-S test on the $\mathrm{log(M_*/M_\odot)}$ distributions of these two samples. The p-value ($\sim$ 0.2) is not sufficiently small to confirm that two distribution are drawn from different underlying distributions. 
We also show the binned median $\mathrm{log(M_*/M_\odot)}$ values of RAGN and the control samples in panel A, which confirms that the control samples are well matched to RAGN. This consistency discounts a possible halo mass effect on their neighbouring galaxies (``Galactic conformity'', e.g.,~\citealt{Phillips2014,Kawinwanichakij2016}), assuming a relation between stellar and halo mass.

As shown in panel B of Figure~\ref{fig:validation}, the two SNG populations reside in the same global environments across the three overdensity bins. We also perform a K-S test on the $\eta$ distributions of these two samples. We obtain a p-value of $\sim$ 0.1, which is larger than the
threshold of rejecting the null hypothesis that the distributions of the two samples are the same. 
The median log($\eta$) values of the two center samples are comparable across the four overdensity bins. The two cumulative distributions are shown in the Figure \ref{fig:control} with a p-value of 0.32 from the K-S test which is not sufficiently small to confirm that the two distribution are drawn from different distributions.  
These tests eliminate the possibility of a global environmental effect on the $\mathrm{f_q}$ of the SNG samples.  

Lastly, we compare the redshift distributions of the two SNG populations as shown in panel C of Figure~\ref{fig:validation}, the binned median z values of RAGN-SNGs versus control-SNGs in each of the three overdensity bins. 
The K-S test of the redshift distributions of all RAGN-SNGs and control-SNGs samples gives a small p-value of $\sim$ 0.01, indicating that they are likely drawn from different distributions. The K-S test of the redshift distributions of the two SNGs samples in the highest overdensity bin also gives a small p-value ($\sim$0.01). However, in the highest overdensity bin, the median z value of RAGN-SNGs and control-SNGs are 0.823$\pm$0.005 and 0.813$^{+0.013}_{-0.060}$, respectively. The two median z values are within 1$\sigma$ difference, which suggests that, at least in the highest overdensity bin, the $f_q$ of neighbouring galaxies is not driven by redshift evolution. 

We have analyzed the representativeness of the control sample and attempted to eliminate all possible factors that might cause the comparisons between the quiescent fraction of RAGN-SNGs and control-SNGs to be biased. Nevertheless, the number of galaxies in these two SNGs sample are small. The control samples are constructed from a limited number of galaxies in the five fields studied here as their stellar mass is, in general, high and such galaxies are relatively rare especially after removing all RAGN hosts from consideration. In order to compensate for this relative lack of galaxies in our primary control samples, we introduce a second control method. 
This method is based on analysis from Lemaux et al. (\textit{submitted}) using 4552 galaxies in 15 ORELSE fields with secure spectroscopic redshifts between 0.55 $\leq$ z $\leq$ 1.40 and with the same stellar mass and colour limits as are applied in this paper. 
Lemaux et al. derived the quiescent fraction relation as a function of redshift, stellar mass, environment $\mathrm{f_{q}(z,  \log(M_{\ast}/M_{\odot}), \log(1+\delta_{gal}))}$ as following the form:
\begin{multline}
\label{eq:global}
$$\mathrm{f_{q}} = (-1.8544\pm0.0002) \times \mathrm{z}^{0.1422\pm0.0007} \\
            + (8.3876\pm0.0001) \times \log(\mathrm{M}_{\ast}/\mathrm{M}_{\odot})^{0.2224\pm0.0007} \\
+ (-11.8321\pm0.002) \times \mathrm{\log(1+\delta_{gal})}^{-0.0075\pm0.0002}$$
\end{multline}
The quiescent fraction calculated from this relation is refereed to as ``global $\mathrm{f_q}$'' hereafter. 
By virtue of the statistical power of thousands of galaxies, we attempt, by adopting this global $\mathrm{f_q}$ as a complimentary control sample, to eliminate any possible residual bias from our comparison to the primary control samples that result from the small number of galaxies present therein. The uncertainty of global $\mathrm{f_q}$ combines the partials of Equation \ref{eq:global} with respect to each variable, using the error on the median of each quantity (redshift, stellar mass, local overdensity), and the uncertainties in the constants shown in Equation \ref{eq:global}. 

The global $\mathrm{f_q}$s of the RAGN-SNGs and control-SNGs are 0.59$\pm$0.02 and 0.60$\pm$0.02, respectively, given their median values of redshift, stellar mass and local overdensity in the highest overdensity bin. 
The ratio between the global $\mathrm{f_q}$ of RAGN-SNGs and that of control-SNGs is 0.98 which is within the 1$\sigma$ uncertainty. Therefore, we exclude bias due to stellar mass, local environment and redshift evolution effects on these two SNGs samples. 
Overall, our measured $\mathrm{f_q}$ offset results in a $\mathrm{f_q}$ ratio of $1.21\pm0.14$ as shown in Figure \ref{fig:result}, a ratio that appears unlikely to be due to the difference of the two SNG populations. 
More noticeable, the measured $\mathrm{f_q}$ of the control-SNGs is only $\sim$1$\sigma$ higher than its global $\mathrm{f_q}$, whereas the measured $\mathrm{f_q}$ of RAGN-SNGs is 5$\sigma$ higher than its global $\mathrm{f_q}$. 
These four $\mathrm{f_q}$s are shown in the Figure \ref{fig:result_comparison}, labeled as the "Highest overdensity bin". 
This result clearly emphasizes the fact that RAGN-SNGs residing in dense environments are more likely to be quenched than their control counterparts in those same environments regardless of the way the control is constructed. Given that the only meaningful difference between the two populations is likely the presence of the RAGN. It is tempting to conclude that this increased quenching is due to the RAGN itself. 
We note that both global $\mathrm{f_q}$s are lower than those in our original SNG samples, which is possibly due to sample variance, as the global $\mathrm{f_q}$s are derived using galaxies in a larger sample less subject to cosmic variance, while $\mathrm{f_q}$s measured in the original RAGN-/control-SNG samples are restricted to the five fields here. Regardless, both control methods have measured or global $\mathrm{f_q}$ values that are consistent at the $<1\sigma$ level. 

\subsection{Effect of Varying $\mathrm{R_{proj}}$}
\label{sec:test_radius}

In this section, we examine the effect of choosing different radii on the result. 
This analysis is motivated by the fact that previous studies looked at the effect of RAGN on neighbouring galaxies selected within smaller radii (e.g., 100 kpc in \citealt{Pace2014}). 
If we use 250 kpc as the radius threshold, there are only 37 galaxies left in the RAGN-SNGs sample in the highest overdensity bin, reduced by $\sim$38.5\%. The quiescent fraction is 0.86 $\pm$ 0.06 for RAGN-SNGs, compared to (0.75$^{+0.07}_{-0.10}$) for the control-SNGs using the same radius threshold. The difference is $\sim$1$\sigma$, which is not as significant as using 500 kpc as the radius threshold. This result could be due to either the large uncertainty derived from the small number of galaxies in these two SNG samples that wash out the signal or the signal does not come mainly from the inner region.
We slightly vary this radial cut to 450 and 550 kpc. The total number of RAGN-SNG is decreased by 12.9\% and increased by 12.4\%, respectively. Similar fractional number of galaxies changed in the highest overdensity bin, with $\mathrm{f_q}$s in these bins equal to 0.81$\pm$0.04 and 0.80$\pm$0.04, respectively. These values are consistent with the values measured using 500 kpc as radial cut.
In addition, we further perform the same analysis on neighbouring galaxies found in an annular radius between 500 kpc to 750 kpc from the RAGN and 100 control samples. We find that quiescent fractions in all local overdensity bins overlap within the uncertainties, and the signal seen in the most dense environment vanishes. This result may indicate that the injection of energy from an RAGN does not affect galaxies on scales larger than 500 kpc. Therefore, we feel confident in our choice to use 500 kpc as the transverse radial search range for neighbouring galaxies. 

We note that the vast majority of RAGN are compact,  beam-size-scale sources, and none of the RAGN extends to the radius we tested, which minimizes the possibility that the neighboring galaxies are directly affected by RAGN radio lobes and the bow shocks driven by the radio lobes on similar scales \citep{Shabala2011}. However, it is possible that the heating through one or multiple cycles of RAGN outbursts could incrementally heat the cluster gas and as gas propagating out to 500 kpc, although we are not able to determine based on the current analysis how many cycles of RAGN activity are needed, or if, indeed, more than a single outburst is necessary to heat the ICM gas up to 500 kpc. 

\subsection{Effect of Radio Luminosity}
\label{sec:test_radio}
In panel C of Figure~\ref{fig:validation}, we show that the median radio luminosity of RAGN studied here increases by 0.48 dex across the three overdensity bins, which indicates a possible correlation between these two properties. 
We apply the Spearman test to assess the correlation between them. The correlation coefficient is 0.36, and the p-value for non-correlation is 0.028. We further test this result adopting a Monte-Carlo simulation where in each iteration we allow the value of luminosity for an individual RAGN to vary based on a Gaussian with a mean and dispersion set to the original $L_{\text{1.4GHz}}$ and its error, respectively. 88\% of Monte-Carlo realizations suggest that there is a correlation (i.e., the Spearman p-value is <0.05). 
All these tests suggest a real positive correlation between local overdensity and luminosity. 
This agrees with the general picture where the relative number of massive galaxies increases with environmental density, especially for quiescent galaxies (see Figure 6 in \citealt{Tomczak2017}). Thus, at fixed accretion mode and accretion rate, massive galaxies which harbor massive black holes would have higher jet power and hence higher radio luminosity. 
The presence of this correlation leaves open the intriguing possibility that the increasing power of RAGN seen in higher density environments lends itself to the quenching of neighbouring galaxies. We will further discuss this potential additive quenching effect in Section \ref{sec:scenario}.

\subsection{Effect of cluster environments}
\label{sec:test_cluster}

The larger quiescent fraction in galaxies surrounding RAGN is only marginally seen in the highest local density region. 10 out of 11 RAGN in that density bin are within the cluster environments as defined in Section \ref{sec:env}\footnote{Though it is possible for a group galaxy to exist at these $\mathrm{log(1+\delta_{gal})}$ and $\eta$ values, we confirmed that none of these RAGN lie within the known group sample in these fields.}, while 6 RAGN are in the virialized core of clusters\footnote{We adopt $\eta \leq 0.1$ as the definition of virialized core region (see \citealt{Noble2013} and reference therein).}. 
One question raised from this result is whether the energy transfer mechanism from the RAGN to its neighbouring galaxies is more efficient in clusters (relative to field or group environments) because the RAGN jet can interact with the ICM, a channel not possible to field/group RAGN. 
There are 24 RAGN that reside in cluster environments and 3 RAGN associated with group environments (i.e., $\eta \leq 2$; see Section \ref{sec:env}). 
Because of the small number of RAGN in group environments, with none of them hosted by the most massive group galaxy, we test the hypothesis using only cluster RAGN. 
There are 106 RAGN-SNGs in the cluster RAGN sub-sample comprising 72\% of the total RAGN-SNGs sample. 68 of these galaxies (96\%) are in the highest overdensity bin. 
We select control samples from a pool of cluster galaxies having $\eta \leq 2$, using the same matching algorithm as described in Section \ref{sec:control}. 
The median number of control-SNGs is $\sim$65\% of the total control-SNGs sample. 
The number of centers in the cluster and their SNGs are shown in the Table \ref{tab:numbers}. 

\begin{figure*}
    \includegraphics[width=\textwidth]{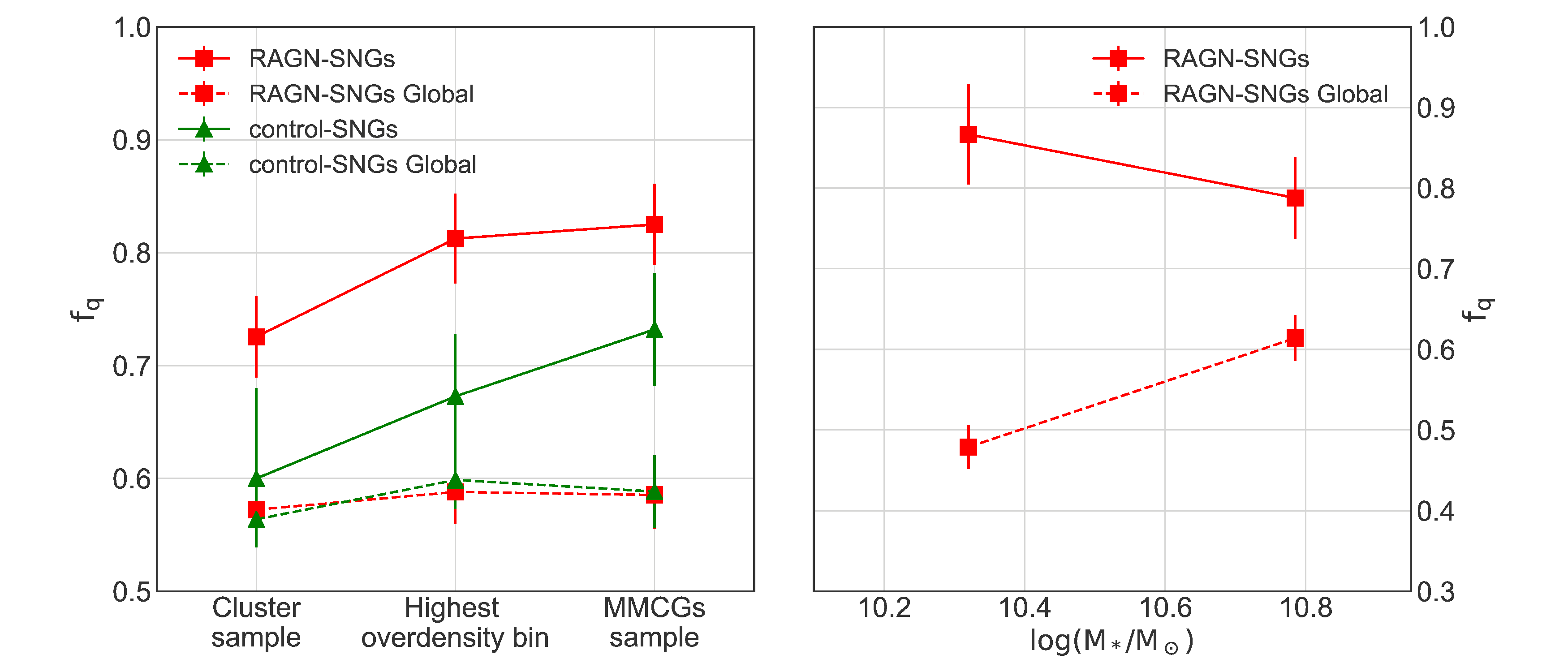}
    \caption{\textit{Left}: The $\mathrm{f_q}$s for RAGN and control samples in three high density environment comparisons. The RAGN-/control-SNGs are samples with their RAGN/control in the highest overdensity bin (1.2 $\mathrm{\leq log(1+\delta_{gal}}) < $ 2.5), the same range as in Figure \ref{fig:result}. The cluster RAGN-/control-SNGs samples are samples with their center RAGN/control having $\eta \leq 2$. The MMCG RAGN-/control-SNGs samples are samples with their centers hosted by the MMCGs. The cluster and MMCGs comparison are further discussed in Section \ref{sec:test_cluster}. The error on RAGN and MMCG control samples are derived from Poissonian statistics and the errorbars on clusters and highest overdensity bin control samples are the 16\% and 84\% value of the 100 re-samplings. \textit{Right:} The quiescent fraction for high/low stellar mass subsamples. The red points are the observed quiescent fraction for high/low stellar mass subsamples separated from RAGN-SNGs in the highest density bin. The $\mathrm{f_q}$s with errors connected by dashed lines in both panels are estimated from the derived quiescent fraction relation presented in Section~\ref{sec:test_comparison}. }
    \label{fig:result_comparison}
    \vspace{-0.4cm}
\end{figure*}

We then compare the $f_q$ of neighbouring galaxies of this RAGN sample with that of the control samples. 
We obtain the $f_q$ of 0.73$\pm$0.04 for RAGN versus 0.60$^{+0.06}_{-0.07}$ for the control samples as shown in the left panel of Figure \ref{fig:result_comparison}, labeled as ``Cluster sample". The error on the control sample value is derived from the variation of 100 control resamplings. 
The offset between the RAGN-SNGs and control-SNGs at 1.8$\sigma$ significant level is consistent with our result in the highest local overdensity bin. However, in the cluster RAGN sample, the RAGN are spread across all three overdensity bins. 
As a comparison, we calculate the global $f_q$ based on the global relation described in Section~\ref{sec:test_comparison} given their median values of stellar mass, redshift and local overdensity. The global $f_q$s are also shown in the left panel of Figure \ref{fig:result_comparison}. The quiescent fraction of the control-SNGs is $\sim1\sigma$ higher than its global $\mathrm{f_q}$, whereas the quiescent fraction of RAGN-SNGs is $4\sigma$ higher than its global $\mathrm{f_q}$. 
We note that both measured $\mathrm{f_q}$s are higher than their global $\mathrm{f_q}$s; nevertheless, as mentioned earlier, the differences between the controls and their corresponding global $\mathrm{f_q}$s are within the 1$\sigma$ uncertainty. 
The 4$\sigma$ offset between the $\mathrm{f_q}$ of RAGN-SNGs in the cluster and its global $\mathrm{f_q}$ tentatively suggests that RAGN within cluster environments might have a larger effect on their neighbouring galaxies, relative to the field or group environment. This large offset may imply that at least some physical mechanism special to the interaction between the RAGN and the cluster environment is capable of affecting the star formation of neighbouring galaxies (see discussion in Section~\ref{sec:discussion}). 

Along this line, studies have suggested that RAGN hosted by the brightest cluster galaxies act as heating agents powerful enough to prevent further cooling of the ICM and regulate density and entropy of the ICM in the cluster environments (e.g., \citealp{Mittal2009, McNamara2007, Best2007}). 
Moreover, other studies suggested that the heating effects of AGN activity from any cluster galaxies, not only those hosted by the Most Massive Cluster Galaxies (MMCGs), might be sufficient to heat the surrounding ICM at larger radii \citep{Nusser2006, Fabian2006}. We use MMCGs to represent the brightest cluster galaxy population in this paper, since there is an enormous overlap between these two types of galaxies in our sample \citep{Rumbaugh2018}. 
To determine whether RAGN hosted by MMCGs have a larger effect on their neighbouring galaxies than MMCGs without RAGN, we implement a comparison of RAGN that are hosted by MMCGs with a control sample. 
There are 5 RAGN hosted by the MMCGs. Here we define the criteria of MMCGs as $\eta \leq 2$, R $\leq$ R$_{200}$, and the galaxy being among the top 3 most massive galaxies in a cluster. Three of these RAGN have 6 companion RAGN within $\mathrm{R_{proj}} \leq$ 500 kpc. Therefore, to consider their feedback together, we categorize these 11 RAGN as the overall RAGN-MMCGs sample. 
Because RAGN-MMCGs are always the 1st or 2nd mass-ranked MMCGs, we use the rest of the 1st and 2nd mass-ranked MMCG candidates as the control-MMCGs. 
The number of RAGN/control-MMCGs and their SNGs are shown in the Table \ref{tab:numbers}. 
The $\mathrm{f_q}$s are 0.83$\pm$0.03 for RAGN-SNGs and 0.70$\pm$0.05 for the control-SNGs and are shown as "MMCGs sample" in the left panel of Figure \ref{fig:result_comparison}. The offset is at a 2.2$\sigma$ significant level, consistent with the offset seen in the cluster comparison and in the highest overdensity bin. 
In addition, we calculate the corresponding global $\mathrm{f_q}$s of RAGN- and control-MMCGs using the relation described in Section~\ref{sec:test_comparison}. These values are shown in the Figure \ref{fig:result_comparison}. The quiescent fractions of SNGs of the RAGN-/control-MMCGs are both much higher than their corresponding global $\mathrm{f_q}$s. However, MMCGs are a special type of galaxy whose effects are not considered in the global $\mathrm{f_q}$ relation. Although we present the global values for completeness, we do not think that a comparison to our second control method is valid in this case. 

Since the MMCG sample is small, to preclude a single MMCG or cluster from dominating the signal, we apply a modified jackknife method. In each iteration, we select 10 out of total 11 galaxies from the MMCG-RAGN/control sample and calculate the $\mathrm{f_q}$ of neighbouring galaxies of the selected centers. The standard deviation on the median $\mathrm{f_q}$ of jackknife re-samplings are 0.001 and 0.004, respectively. The small deviations indicate that the results are not dominated by a single centre galaxy. 
We notice that 4 out of 11 RAGN and 32 out of total 67 RAGN-SNGs are associated with the most massive cluster in RXJ1821. 
Additionally, the RXJ1821 cluster is unique in the sample studied here in terms of its mass and compactness \citep{Rumbaugh2012, Shen2017}.
We attempted an analysis that excludes all SNGs in this cluster, finding no significant difference between quiescent fractions of SNGs of MMCG-RAGN and that of MMCG-control. However, the small sample size of the remaining galaxy population essentially precludes a significant result. 
Furthermore, we attempted to compare these 4 RAGN and their associated RAGN-SNGs in the RXJ1821 cluster to the rest of its cluster members. The $\mathrm{f_q}$ of the RAGN-SNGs sub-sample is higher than that of the other cluster members at $>3\sigma$, even when we match the median $\eta$ value of the two sub-samples. However, we only have this one cluster that has mutiple RAGN in the cluster centre so we cannot draw any firm conclusions. We will explore in future work, using the power of the full ORELSE sample, whether such clusters are primarily responsible for driving the elevated $\mathrm{f_q}$ seen here.
As we mentioned in Section \ref{sec:agn_sample} and shown in Figure \ref{fig:cutout}, there is one extended RAGN source hosted by the MMCG in the main RXJ1821 cluster. This RAGN is not coincident with the X-ray center, but is actually $\sim$700 kpc away. We test the impact of this special case by removing the 9 SNGs of this RAGN from the RAGN-SNGs sample and re-calculating the fq. None of the results change meaningfully. 

Radio AGN in clusters are known to be associated with strong cool-core clusters, clusters which are typically in a relaxed state (e.g. \citealp{Mittal2009}, \citealp{Cavaliere2016}). Relaxed clusters, in general, might be expected to have higher quiescent fractions (e.g. \citealp{Lemaux2012}). These cool-core clusters, however, are probably not prevalent in our sample and, regardless, our data do not have the ability to discern such phenomenon. In addition, although RAGN are preferentially found in regions of higher global density \citep{Shen2017}, they are not typically found in the centre of the clusters that are studied here. Therefore, they are likely not the type of RAGN seen in cool-core clusters. However, a concern still remains that, if our RAGN sample reside in the clusters that have a higher average $\mathrm{f_q}$ than that of clusters which host the control samples, then our result could be biased in exactly the direction we are claiming a potential signal. As we show here, however, this bias is almost certainly not present in our sample. From the analysis presented in \citealp{Rumbaugh2018}, we confirm that RAGN in the cluster and MMCGs sub-samples reside in clusters having various $\mathrm{f_q}$s spanning from 0.38 to 0.78. Furthermore, we see no evidence of correlation between the number of RAGN and the fraction of quiescent members in the same cluster. These results imply that the elevated $\mathrm{f_q}$s seen in the cluster and MMCG comparisons are not driven by other unrelated processes in their host clusters. 
As emphasized in the left panel of Figure~\ref{fig:result_comparison}, no matter what comparison we made, we have $\sim$2$\sigma$ significant between $\mathrm{f_q}$s for RAGN-SNG and that for all other control-SNG samples. We will further discuss the quenching effect of RAGN in clusters and those being hosted by MMCGs in Section \ref{sec:scenario}. 

\subsection{Effect on low stellar mass SNGs}
\label{sec:test_lowmass}

Previous studies suggested that both RAGN feedback and cluster physical mechanisms, such as ram pressure stripping, might have larger effects on low stellar mass galaxies (e.g., ~\citealt{Shabala2011, Bahe2015}). 
Therefore, we separate RAGN-SNGs in the highest density bin, as shown in Figure \ref{fig:result}, into two sub-samples: low stellar mass ($\mathrm{M_* \leq 10^{10.5}M_{\odot}}$) and high stellar mass ($\mathrm{M_* \ge 10^{10.5}M_{\odot}}$) galaxies. 
We find that the low stellar mass sub-sample has a higher quiescent fraction ($\mathrm{f_q} = 0.87 \pm 0.06$) than the high stellar mass sub-sample ($\mathrm{f_q} = 0.79 \pm 0.05$), as shown in the right panel of Figure \ref{fig:result_comparison}. 

As discussed in Section~\ref{sec:test_comparison}, studies have found that the quiescent fraction of galaxies depends on various parameters,  e.g., stellar mass, environment, and redshift \citep{Muzzin2013, Tomczak2014, Cooke2016}. Thus, to make a fair comparison between the two stellar mass sub-samples, we should apply corrections to their quiescent fraction based on the expected values at their respective average $\mathrm{M_{\ast}}$, z, and $\mathrm{\log(1+\delta_{gal})}$ values.  
Specifically, the quiescent fraction has been found to be strongly tied to each of these three parameters using $\sim$4500 spectroscopically confirmed galaxies across 15 ORELSE fields (see Lemaux et al. \textit{submitted}). Lemaux et al. found that the quiescent fraction, on average, smoothly decreases with decreasing stellar mass at all redshifts for galaxies residing in all environments. 
Based on the global relation presented in Section~\ref{sec:test_comparison}, the global $\mathrm{f_q}$s of the lower and higher stellar mass sub-samples are 0.48 $\pm$ 0.02 and 0.61 $\pm$ 0.02, respectively, given the median stellar masses (10.32 vs 10.79), redshifts (0.821 vs 0.859) and local overdensity (1.22 vs 1.33) of these RAGN-SNG sub-samples. These global $\mathrm{f_q}$ values are shown in the right panel of Figure \ref{fig:result_comparison}, along with two measured $\mathrm{f_q}$s for low and high stellar mass RAGN-SNG sub-samples. The offset between global and measured $\mathrm{f_q}$s are 6$\sigma$ and 3$\sigma$ for the low and high stellar mass RAGN-SNG sub-samples, respectively. This result clearly shows that RAGN-SNGs in dense environments are more likely to be quenched than general ORELSE galaxies having similar stellar mass and redshift and residing in similar environments. 
Furthermore, assuming everything else being equal, the $f_q$ ratio should be 0.78 between the low stellar mass and high stellar mass sub-samples. However, this is contrary to what we find for the two sub-samples studied here, as the $f_q$ ratio is 1.10 $\pm$ 0.10. The difference between our measurement and the global ratio has a $\sim$3.2$\sigma$ significance, which suggests that the feedback from nearby RAGN is more effective on low stellar mass neighbouring galaxies. We note that we do not compare to control/control-SNG, due to the small number of control-SNGs in low/high stellar mass sub-samples. Instead, the global $\mathrm{f_q}$ which is derived from the full ORELSE sample gives a more reliable comparison in this case.  

\section{DISCUSSION AND SUMMARY}
\label{sec:discussion}
Numerous studies have shown that feedback from RAGN can affect the star formation of their host galaxies and has the capability of heating the surrounding ICM (e.g., \citealt{Birzan2004, Voit2005, McNamara2012}). 
Previous attempts to study the possibility of RAGN quenching of their neighbouring galaxies have been unable to reveal any signs of this effect except under special circumstances. However, in this paper, we searched for the signature of this effect to larger radius and across a wider dynamical range of environments, using 57 RAGN and their neighbouring galaxies ($\mathrm{R_{proj}}\leq$500 kpc) at intermediate redshifts ($0.55 \leq z \leq 1.3$) selected from five fields in the ORELSE survey. 
To isolate the effect of RAGN, we selected 100 control galaxies which match the colour, stellar mass and local overdensity of RAGN (see Section \ref{sec:control}) and obtained neighbouring galaxies of the control sample (see Section \ref{sec:Ngals}). We calculated the fractions of quiescent galaxies for RAGN-SNGs and control-SNGs by binning the two samples into four local overdensity bins and found a marginal (2$\sigma$) increase in $\mathrm{f_q}$ of RAGN-SNGs compared to control-SNGs at the highest densities of $\mathrm{\log(1+\delta_{gal})} \geq 1.2$, but in no other local environments. 

To confirm the validity of the comparisons made, we ran diagnostic tests in Section \ref{sec:test_comparison} and Section \ref{sec:test_radius}. We first examined whether this possible difference is due to the intrinsic differences between RAGN-SNGs and control-SNGs which may bias the $\mathrm{f_q}$ based on the stellar mass, environment, and/or redshift. We exclude these possible effects on $\mathrm{f_q}$ of the two SNG samples and even the RAGN and control samples. We then use different radii to select neighbouring galaxies and found that the largest difference between the RAGN-SNG and control-SNG samples is seen using 500 kpc as $\mathrm{R_{proj}}$. Using smaller radial range reduces the sample size of SNGs, which increases the uncertainty of $\mathrm{f_q}$. In addition, small variations on the radius cut ($\pm$ 50 kpc) do not affect our result. Further, we found the signal vanished using a larger radial range, which could be explained by the fact that RAGN might not be capable of affecting their neighbouring galaxies to such large distances. 

After eliminating possible effects from factors other than the RAGN, we searched for the origin of this difference from radio power in Section \ref{sec:test_radio} and cluster environments in Section \ref{sec:test_cluster}. 
We found that the median values of radio power of RAGN increases with increasing local overdensity, which suggests that the increased radio power of RAGN in high-density environments might be one of the potential factors that enhances the quenching of RAGN-SNGs. 
Because the observed difference in $\mathrm{f_q}$ occurs at very low log($\eta$) and because $\sim$50\% of RAGN and 62\% of SNGs in our sample are within the cluster environments, we examined whether RAGN combined with the cluster environment may cause the larger quenching effect in the highest overdensity bin.  
We performed a comparison of RAGN in clusters and those hosted by MMCGs with a matched cluster/MMCG control sample and found a 2$\sigma$ increase as in our previous result. 
When comparing $\mathrm{f_q}$s for RAGN- and control-SNGs in clusters and the highest overdensity bin to their $\mathrm{f_q}$s from the derived global relation between quiescent fraction and stellar mass, redshift, and local overdensity from Lemaux et al. (\textit{submitted}), a more significant offset is observed between RAGN-SNGs and its global $\mathrm{f_q}$, compared to the $\mathrm{f_q}$ of the primary control-SNGs and its global $\mathrm{f_q}$. 
In addition, we find a 6$\sigma$ offset relative to its global $\mathrm{f_q}$ for the lower stellar mass RAGN-SNGs versus a $3\sigma$ offset for the higher stellar mass subsample, as well as the inverse relation of $\mathrm{f_q}$ for RAGN-SNGs depending on stellar mass compared to the global values was found. 
Both of these results imply that the RAGN have a larger effect on low mass galaxies, as might be expected. 

\subsection{Possible Interpretations for RAGN Induced Quenching of neighbouring Galaxies}
\label{sec:scenario}
Emboldened by the observation of a persistently higher incidence of quiescence of galaxies in close proximity of a RAGN, we propose here possible interpretations of why RAGN residing in clusters might act to decrease the capability of surrounding galaxies to form stars. Because of their location within an overarching diffuse medium where RAGN are thought to be able to heat the surrounding ICM and to potentially enhance the physical mechanisms which remove galaxy gas, RAGN could have the consequence of quenching star formation in neighbouring galaxies.

Many studies of clusters at low redshift using both X-ray and radio observations have revealed that AGN deposit energy via jets and bubbles and moves ICM gas to outer cluster regions via weak shocks, with the latter mechanism acting on a larger scale ($\sim$300 kpc). Recent observations of very large-scale and diffuse radio structures around 3C 31 and earlier observations of M87 suggest that a large-scale ($\sim$200 kpc) heat deposition may be taking place \citep{Hardcastle2002, Owen2000}. \citet{Ma2011} studied ICM atmospheric heating via RAGN jets in the redshift range $0.1\leq z\leq 0.6$. They found that those RAGN residing within a projected radius of 250 kpc from the cluster center are able to heat gas in the ICM on order of $\sim$0.2 keV per particle within R$_{500}$, which corresponds to $\sim$700 kpc for clusters in these five fields. 
In our sample, the average radio power $\text{P}_{\text{radio}}$ at 1.4GHz is $10^{24.2} \mathrm{W\ Hz}^{-1}$ in the highest overdensity bin. Following the $\mathrm{P_{jet}-P_{radio}}$ scaling relations in \citet{Cavagnolo2010}, we obtain a jet power ($\mathrm{P_{jet}}$) of $\sim10^{44}$ erg/s, similar to the average RAGN jet power studied in \citet{Ma2011}. 
We assume this $\sim$0.2 keV increase of $\mathrm{T_{ICM}}$ up to R$_{500}$ in our clusters. 
Given that the median $\mathrm{T_{ICM}}$ of our cluster sample is 3.7 KeV, as measured within core radii of 180 kpc from Chandra observations (see more details in \citealt{Rumbaugh2018}), the corresponding increase in the median value of $\mathrm{T_{ICM}}$ is 5\% with a range from 2\% to 25\%. 
The estimated mass loss rate due to hydrodynamic interactions (i.e., viscous striping and thermal evaporation) taking place between the galaxy's ISM and the ICM is $\dot{\text{M}} \propto \text{T}_{\text{ICM}}^{2.5}$ (see \citealt{Boselli2006} for reference). 
Therefore, given the estimated $\text{T}_{\text{ICM}}$ increase, $\dot{\text{M}}$ could increase by $\sim$ 15\%, with the range of 6\% to 75\%. 
In summary, this simplified estimation of the temperature change and its consequence on mass loss rate supports our scenario that the heating mechanism of RAGN could remove gas from a galaxy that would be available for star formation and consequently of quench star formation in neighbouring galaxies. 
To this end, some simulations of RAGN shown that multiple duty cycles lead to the depositing of considerable energy to the ICM (e.g. \citealp{Voit2005}), though it is unclear if any of the RAGN in our sample are comparable to the phenomenon simulated (see discussion in Section~\ref{sec:test_cluster}.)

Here, we further discuss three factors which might affect our scenario: radio power, stellar mass of neighbouring galaxies, and effects related to RAGN being hosted by MMCGs. The radio power $\text{P}_{\text{radio}}$ at 1.4GHz of our RAGN sample spans $10^{23.22} \sim 10^{25.13} \mathrm{W\ Hz}^{-1}$ and, as shown in the panel C of Figure \ref{fig:result}, the median values of radio power increase by 0.48 dex across the three overdensity bins. Assuming the same $\text{P}_{\text{jet}}-\text{P}_{\text{radio}}$ scaling relation as above, the corresponding $\text{P}_{\text{jet}}$ range across these bins is $10^{43.33} \sim 10^{45.75}$ erg/s. This estimated range results in a $\sim$2.5 dex change from the low- to the high-density environments in $\text{P}_{\text{jet}}$. However, there is an extremely large scatter in this scaling relation ($\sim$3.5 dex). As such, we are not able to definitively claim that the increase in radio power observed from low- to high-density environments contributes to the possible quenching effect. 

In Section \ref{sec:test_lowmass}, the lower stellar mass RAGN-SNGs were shown to have a higher quiescent fraction both relative to their higher stellar mass RAGN-SNGs and relative to the global value derived from the full ORELSE sample. This increased $\mathrm{f_q}$ for lower stellar mass galaxies in the densest regions of massive clusters evokes thoughts of ram pressure stripping, which is more efficient in stripping gas in the lower stellar mass galaxies mostly due to their shallower potential well \citep{Boselli2006}. 
Therefore, the more pronounced effect for lower stellar mass neighbouring galaxies seen here could be explained by the additive effect of heating by RAGN, which would further weaken an already shallow potential well, so that ram pressure stripping effects could excise most of the in situ gas. In galaxies with a steeper potential well, i.e., higher stellar mass galaxies, the difference in temperature induced by the RAGN may not be enough to make an appreciable difference in ram pressure stripping effectiveness. 

In case of RAGN hosted by MMCGs, we found an offset between RAGN-SNGs and control-SNGs at a 2$\sigma$ significance level, which is similar to the offset seen in the cluster comparison and in the highest overdensity bin. 
Many studies have demonstrated that RAGN within the cooling radius of clusters (i.e. the radius within which the cooling time is less than the Hubble time) could regulate heat, density and entropy of the ICM (e.g., \citealt{Mittal2009, McNamara2007}), with such heating exceeding that of all other RAGN in a given cluster combined \citep{Best2007}. However, the fact that the small sample size of the MMCG population essentially precludes a more significant result. 
In addition, these RAGN typically have nearby RAGN in our sample, which appear to be another factor that can enhance the quenching effect. This result is in line with studies that suggested that the heating effects of AGN activity from all cluster galaxies might be a solution to insufficient heating of the ICM at larger radius ($>$ R$_{200}$) of the cluster center \citep{Nusser2006, Fabian2006}. We will explore in future work, using the power of the full ORELSE sample, whether RAGN being hosted by MMCGs are responsible for driving the elevated $\mathrm{f_q}$ seen here. 

As for non-cluster environments, \citet{Giodini2010} studied the mechanical energy output from 16 group RAGN up to z$\sim$1. They found that the energy released by RAGN is larger than gravitational binding energy of the intragroup medium. They suggested that gas in the group has been removed by RAGN. 
In the Local Group, an additional RAGN feedback mechanism is found via strong shocks by powerful FR-II type radio sources \citep{Worrall2012}, in line with the feedback mechanism suggested by \citet{Shabala2011}. 
Unfortunately, we could not test in group environments, since our current sample only contains 3 RAGN in known galaxy groups, and none of them reside in the center of groups in the ORELSE fields being studied in this paper. 
On the other hand, RAGN in field environments may not be able to transfer their energy output effectively to neighbouring galaxies. This could be explained due to the lack of a medium necessary to couple the mechanical energy of the RAGN jets. Additionally, in situ HI/HII gas in neighbouring galaxies is transparent to the RAGN jets (i.e. jet are not capable of interacting with galaxies in situ HI/HII gas; \citealp{McNamara2012}). Therefore, we would not expect significant effects in the field environments as we have found in the cluster environments. 

There still remain, however, some caveats in our analyses. Given the size of our total RAGN sample and RAGN-cluster sample, we are not able to definitively confirm our result at least for some aspects of this analysis. In addition, the spatial selection of our spectroscopy could affect our result. Though we attempt to mitigate the effects that such sampling could have on our results by selecting 100 different control samples, it is at least conceivable that some differential bias between the RAGN-SNGs and control-SNGs persists. This is, additionally, a bias which may be compounded for RAGN which emit jets with smaller opening angles. The full ORELSE sample, which is expected to provide a RAGN sample which exceeds the current sample by a factor of approximately three, should be efficient to settle these issues and to definitively determine whether the RAGN-induced quenching suggested here is real or not.

\section*{Acknowledgements}

This material is based upon work supported by the National Science Foundation under Grant No. 1411943. Part of the work presented herein is supported by NASA Grant Number NNX15AK92G. 
This study is based on data taken with the Karl G. Jansky Very Large Array which is operated by the National Radio Astronomy Observatory. The National Radio Astronomy Observatory is a facility of the National Science Foundation operated under cooperative agreement by Associated Universities, Inc. 
This work is based, in part, on data collected at the Subaru Telescope and obtained from the SMOKA, which is operated by the Astronomy Data centre, National Astronomical Observatory of Japan; observations made with the Spitzer Space Telescope, which is operated by the Jet Propulsion Laboratory, California Institute of Technology under a contract with NASA; and data collected at UKIRT which is supported by NASA and operated under an agreement among the University of Hawaii, the University of Arizona, and Lockheed Martin Advanced Technology centre; operations are enabled through the cooperation of the East Asian Observatory. When the data reported here were acquired, UKIRT was operated by the Joint Astronomy Centre on behalf of the Science and Technology Facilities Council of the U.K. 
This study is also based, in part, on observations obtained with WIRCam, a joint project of CFHT, Taiwan, Korea, Canada, France, and the Canada-France- Hawaii Telescope which is operated by the National Research Council (NRC) of Canada, the Institut National des Sciences de l'Univers of the Centre National de la Recherche Scientifique of France, and the University of Hawai'i. The scientific results reported in this article are based in part on observations made by the Chandra X-ray Observatory and data obtained from the Chandra Data Archive. 
The spectrographic data presented herein were obtained at the W.M. Keck Observatory, which is operated as a scientific partnership among the California Institute of Technology, the University of California, and the National Aeronautics and Space Administration. The Observatory was made possible by the generous financial support of the W.M. Keck Foundation. 
We wish to thank the indigenous Hawaiian community for allowing us to be guests on their sacred mountain, a privilege, without with, this work would not have been possible. We are most fortunate to be able to conduct observations from this site.

\bibliographystyle{mnras}
\bibliography{reference}  




\bsp	
\label{lastpage}
\end{document}